\documentclass[fleqn,conference]{IEEEtran}
\IEEEoverridecommandlockouts
\usepackage[utf8]{inputenc}
\usepackage{cite}
\usepackage{url}
\usepackage{amsmath,amssymb,amsfonts}
\usepackage[ruled,vlined,linesnumbered]{algorithm2e}
\usepackage{algorithmic}
\usepackage{graphicx}
\usepackage{textcomp}
\usepackage[x11names,dvipsnames,table]{xcolor}
\usepackage{comment}
\def\BibTeX{{\rm B\kern-.05em{\sc i\kern-.025em b}\kern-.08em
    T\kern-.1667em\lower.7ex\hbox{E}\kern-.125emX}}
\usepackage{multirow}
\usepackage{tikz}
\usetikzlibrary{decorations.pathreplacing}
\usetikzlibrary{matrix}
\usepackage{pgfplots}
\pgfplotsset{compat=1.18}
\pgfplotsset{every axis legend/.append style={font=\tiny}}

\usepackage[font={small}]{caption, subfig}
\DeclareMathSizes{10}{9}{6}{5}

\usepackage[nonumberlist]{glossaries}
\usepackage{mdframed} 
\usepackage{multicol} 
\usepackage{enumitem} 

\definecolor{rangeA1}{RGB}{255,255,179} 
\definecolor{rangeA2}{RGB}{255,255,128} 
\definecolor{rangeA3}{RGB}{255,255,77}  
\definecolor{rangeA4}{RGB}{255,255,0}   
\definecolor{rangeA5}{RGB}{230,230,0}   
\definecolor{rangeA6}{RGB}{204,204,0}   

\definecolor{rangeB1}{RGB}{255,230,179} 
\definecolor{rangeB2}{RGB}{255,204,128} 
\definecolor{rangeB3}{RGB}{255,179,77}  
\definecolor{rangeB4}{RGB}{255,153,0}   
\definecolor{rangeB5}{RGB}{230,138,0}   
\definecolor{rangeB6}{RGB}{204,122,0}   

\definecolor{rangeC1}{RGB}{255,179,179} 
\definecolor{rangeC2}{RGB}{255,128,128} 
\definecolor{rangeC3}{RGB}{255,77,77}   
\definecolor{rangeC4}{RGB}{255,0,0}     
\definecolor{rangeC5}{RGB}{204,0,0}     
\definecolor{rangeC6}{RGB}{153,0,0}     

\definecolor{rangeD1}{RGB}{179,255,179} 
\definecolor{rangeD2}{RGB}{128,255,128} 
\definecolor{rangeD3}{RGB}{77,255,77}   
\definecolor{rangeD4}{RGB}{0,255,0}     
\definecolor{rangeD5}{RGB}{0,204,0}     
\definecolor{rangeD6}{RGB}{0,153,0}     

\definecolor{rangeE1}{RGB}{179,179,255} 
\definecolor{rangeE2}{RGB}{128,128,255} 
\definecolor{rangeE3}{RGB}{77,77,255}   
\definecolor{rangeE4}{RGB}{0,0,255}     
\definecolor{rangeE5}{RGB}{0,0,204}     
\definecolor{rangeE6}{RGB}{0,0,153}     

\definecolor{exact}{RGB}{0,0,0} 

\newif\ifcomm
\commtrue 
\ifcomm
\else
\commfalse
\fi
\ifcomm
\newcommand\ak[1]{\textcolor{orange}{AK: #1}}
\newcommand\ogi[2]{\textcolor{magenta}{Ogi: #1}}

\else
\newcommand\ak[1]{}
\newcommand\ogi[2]{}
\fi

\newglossary{printedglossary}{not}{ntn}{Nomenclature}
\newglossary{internal}{sym}{sbl}{internal wont be printed}

\makeglossaries

\newglossarystyle{customlist}{%
  \setglossarystyle{list}%
}

\setglossarystyle{customlist}

\newglossaryentry{networkgraph}
{ type=printedglossary, name={\ensuremath{G}}, sort={01}, 
description={Directed network graph} } 

\newglossaryentry{node}
{ type=printedglossary, name={\ensuremath{n}}, sort={02}, description={Node} }

\newglossaryentry{set_of_nodes}
{ type=printedglossary, name={\ensuremath{N}}, sort={03}, description={Set of nodes} }

\newglossaryentry{wired_bridge}
{ type=printedglossary, name={\ensuremath{dbr}}, sort={04},
    description={Wired bridge} }

\newglossaryentry{wireless_bridge}
{ type=printedglossary, name={\ensuremath{sbr}}, sort={05},
    description={Wireless bridge} }

\newglossaryentry{wireless_station}
{ type=printedglossary, name={\ensuremath{ses}}, sort={06},
    description={Wireless end station} }

\newglossaryentry{wired_station}
{ type=printedglossary, name={\ensuremath{des}}, sort={07},
    description={Wired end station} }

\newglossaryentry{set_of_links}
{ type=printedglossary, name={\ensuremath{L}}, sort={08}, description={Set of links} }

\newglossaryentry{wiredlink}
{ type=printedglossary, name={$\ensuremath{l}^{\ensuremath{wired}}$}, sort={09},
    description={Wired link} }

    \newglossaryentry{wirelesslink}
{ type=printedglossary, name={$\ensuremath{l}^{\ensuremath{wireless}}$}, sort={10},
    description={Wireless link} }

\newglossaryentry{luplink}
{ type=printedglossary, name={$\ensuremath{l}^{\ensuremath{wireless}}_{\ensuremath{uplink}}$}, sort={11},
    description={Wireless uplink} }

\newglossaryentry{ldownlink}
{ type=printedglossary, name={$\ensuremath{l}^{\ensuremath{wireless}}_{\ensuremath{downlink}}$}, sort={12},
    description={Wireless downlink} } 

\newglossaryentry{func_ul_n}{
    type=printedglossary, name={\ensuremath{ul(\gls{node})}}, sort={13},
    description={Function returns \gls{luplink} for \gls{node}}}

\newglossaryentry{func_dl_n}{
    type=printedglossary, name={\ensuremath{dl(\gls{node})}}, sort={14},
    description={Function returns \gls{ldownlink} for \gls{node}}}

\newglossaryentry{stream}
{ type=printedglossary, name={\ensuremath{s}}, sort={15}, description={Data stream index} }

\newglossaryentry{set_of_streams}
{ type=printedglossary, name={\ensuremath{S}}, sort={16}, description={Set of data streams} }

\newglossaryentry{port}
{ type=printedglossary, name={\ensuremath{p}}, sort={17}, description={Egress port name} }

\newglossaryentry{set_of_ports}
{ type=printedglossary, name={\ensuremath{P}}, sort={18}, description={Set of egress ports} }

\newglossaryentry{period_s}
{ type=printedglossary, name={\ensuremath{C_{\gls{stream}}}}, sort={19},
description={Cycle period of \gls{stream}} }

\newglossaryentry{lcm}
{ type=printedglossary, name={\ensuremath{LCM_{\gls{set_of_streams}}}}, sort={20}, description={Least common multipliers of all \gls{period_s}}}

\newglossaryentry{repetition}
{ type=printedglossary, name={\ensuremath{r}}, sort={21}, description={Repetition instance index of \gls{stream}} }

\newglossaryentry{path}
{ type=printedglossary, name={\ensuremath{u}}, sort={22}, description={Path index}}

\newglossaryentry{set_of_paths}
{ type=printedglossary, name={\ensuremath{U_{\gls{stream}}}}, sort={23}, description={Function returns set of paths for \gls{stream}}}

\newglossaryentry{r_sp}
{ type=printedglossary, name={\ensuremath{\Gamma}}, sort={24}, description={Robustness level} }

\newglossaryentry{ft_kp}
{ type=printedglossary, name={\ensuremath{ft_s(k,\gls{port})}}, sort={25}, description={Time needed for frame at \gls{port}}}

\newglossaryentry{fd_kp}
{ type=printedglossary, name={\ensuremath{fd_s(k,\gls{port})}}, sort={26}, description={Maximum deviation for frame at \gls{port}}}

\newglossaryentry{func_tsp}
{ type=printedglossary, name={\ensuremath{t(\gls{stream},\gls{port})}}, sort={27},
description={Cumulative time for \gls{stream}}}

\newglossaryentry{func_dsp}
{ type=printedglossary, name={\ensuremath{d(\gls{stream},\gls{port})}}, sort={28},
description={Cumulative  deviation for \gls{stream}} }

\newglossaryentry{max_latency}
{ type=printedglossary, name={\ensuremath{max\_latency_{\gls{stream}}}}, sort={29}, description={Maximum latency of \gls{stream}} }

\newglossaryentry{max_jitter}
{ type=printedglossary, name={\ensuremath{max\_jitter_{\gls{stream}}}}, sort={30}, description={Maximum jitter of \gls{stream}} }

\newglossaryentry{dv_xrusp}
{ type=printedglossary,
name={\ensuremath{x_{\gls{stream},\gls{port}}^{\gls{repetition}, \gls{path}}}}, sort={31},
description={Variable, time \gls{port} opens} }

\newglossaryentry{dv_as}
{ type=printedglossary,
name={\ensuremath{a_{\gls{stream}}}}, sort={32},
description={Variable, \gls{stream} scheduled or not} }

\newglossaryentry{dv_zs}
{ type=printedglossary,
name={\ensuremath{z_{\gls{stream}, \gls{path}}}}, sort={33},
description={Variable, \gls{stream}, \gls{path} scheduled or not} }

\newglossaryentry{dv_y}
{ type=printedglossary,
name={\ensuremath{y_{\gls{stream},\gls{repetition}, \gls{path},\gls{port}}^{\gls{stream_2},\gls{repetition_2},\gls{path_2}}}},
sort={34}, description={Helper variable, \gls{port} no-overlap}}

\newglossaryentry{dv_yf}
{ type=printedglossary,
name={\ensuremath{y_{\gls{stream},\gls{repetition}, \gls{path},\gls{port}}^{f}}},
sort={34}, description={Helper variable, \gls{port} no-overlap}}

\newglossaryentry{dv_w}
{ type=printedglossary,
name={\ensuremath{w_{(\gls{stream},\gls{repetition}, \gls{path},\gls{port})}^{(\gls{stream_2},\gls{repetition_2},\gls{path_2},\gls{port_2})}}},
sort={35}, description={Helper variable, \gls{luplink}, \gls{ldownlink}}}

\newglossaryentry{func_Nsp}
{ type=printedglossary, name={\ensuremath{next\_port_{\gls{stream},\gls{path},\gls{port}}}}, sort={36},
description={Function, returns next port}}

\newglossaryentry{func_ns_sp}
{ type=printedglossary, name={\ensuremath{next(\gls{r_sp}, \gls{stream}, \gls{path}, \gls{port}, \gls{port_2})}}, sort={37},
description={Function, returns next port time}}

\newglossaryentry{func_Rs}
{ type=printedglossary, name={\ensuremath{R_{\gls{stream}}}}, sort={39},
description={Function, returns set of \gls{repetition}} }

\newglossaryentry{func_Ps}
{ type=printedglossary, name={\ensuremath{P_{\gls{stream}, \gls{path}}}}, sort={40},
description={Function, returns of ports at path} }

\newglossaryentry{fixed_start}
{ type=printedglossary, name={\ensuremath{fixed\_start_{f}}}, sort={41}, description={Fixed block start} }

\newglossaryentry{fixed_end}
{ type=printedglossary, name={\ensuremath{fixed\_end_{f}}}, sort={42}, description={Fixed block end} }

\newglossaryentry{fixed_set}
{ type=printedglossary, name={\ensuremath{fixed\_set(\gls{stream}, \gls{port})}}, sort={43}, description={Function, returns fixed for port}}

\newglossaryentry{bigm}
{ type=printedglossary, name={\ensuremath{M}}, sort={45}, description={Big enough
number} }

\newglossaryentry{path_2}
{ type=internal, name={\ensuremath{u'}}, sort={01}, description={Other path}}
\newglossaryentry{set_of_paths_2}
{ type=internal, name={\ensuremath{U_\gls{stream_2}}}, sort={01}, description={Returns feasible set of paths for \gls{stream_2}}}

\newglossaryentry{port_2}
{ type=internal, name={\ensuremath{p'}}, sort={02}, description={Egress port} }
\newglossaryentry{dv_zs2}
{ type=internal,
name={\ensuremath{z_{\gls{stream_2}, \gls{path_2}}}}, sort={04},
description={Decision variable stream \gls{stream_2}, \gls{path_2} scheduled or not} }

\newglossaryentry{dv_x1sp}
{ type=internal, name={\ensuremath{x_{\gls{stream},\gls{port}}^{1, \gls{path}}}}, sort={04},
description={Time stream \gls{stream} leaves port \gls{port} for repetition 1} }

\newglossaryentry{dv_xrusp_2}
{ type=internal,
name={\ensuremath{x_{\gls{stream_2},\gls{port}}^{\gls{repetition_2}, \gls{path_2}}}},
sort={05}, description={Time stream \gls{stream_2} leaves port \gls{port} for
repetition \gls{repetition_2}} }

\newglossaryentry{dv_xrusp_2w}
{ type=internal,
name={\ensuremath{x_{\gls{stream_2},\gls{port_2}}^{\gls{repetition_2}, \gls{path_2}}}},
sort={05}, description={Time stream \gls{stream_2} leaves port \gls{port} for
repetition \gls{repetition_2}} }

\newglossaryentry{dv_xrusp2}
{ type=internal,
name={\ensuremath{x_{\gls{stream},\gls{port_2}}^{\gls{repetition}, \gls{path}}}},
sort={05}, description={Time stream \gls{stream} leaves port \gls{port_2} for
repetition \gls{repetition}} }

\newglossaryentry{dv_xr2usp}
{ type=internal,
name={\ensuremath{x_{\gls{stream},\gls{port}}^{\gls{repetition_2}, \gls{path}}}},
sort={05}, description={Time stream \gls{stream} leaves port \gls{port} for
repetition \gls{repetition_2}} }

\newglossaryentry{stream_2}
{ type=internal, name={\ensuremath{s'}}, sort={08}, description={Other stream} }

\newglossaryentry{stream_3}
{ type=internal, name={\ensuremath{s''}}, sort={08}, description={Other stream} }

\newglossaryentry{repetition_2}
{ type=internal, name={\ensuremath{r'}}, sort={09}, description={Other streams
repetition} }
\newglossaryentry{func_Rs_2}
{ type=internal, name={\ensuremath{R_{\gls{stream_2}}}}, sort={12},
description={Function returns set of repetition \gls{repetition} given stream
\gls{stream_2}} }
\newglossaryentry{func_P_inv_p}
{ type=internal, name={\ensuremath{P'_{\gls{port}}}}, sort={14},
description={Function returns set of streams \gls{stream_2} given port
\gls{port}} }
\newglossaryentry{func_ds2p}
{ type=internal, name={\ensuremath{d(\gls{stream_2},\gls{port})}}, sort={15},
description={Expected maximum throughput time deviation for \gls{stream} at port
\gls{port} , where other stream should not be assigned.} }

\newglossaryentry{func_ds2pw}
{ type=internal, name={\ensuremath{d(\gls{stream_2},\gls{port_2})}}, sort={15},
description={Expected maximum throughput time deviation for \gls{stream} at port
\gls{port} , where other stream should not be assigned.} }

\newglossaryentry{func_dsp2}
{ type=internal, name={\ensuremath{d(\gls{stream},\gls{port_2})}},
sort={16}, description={Expected maximum throughput time deviation for
\gls{stream} at port where port is previous of port \gls{port}. If previous port
is not exist, taken as 0.} }

\newglossaryentry{r_s2p}
{ type=internal, name={\ensuremath{\Gamma}}, sort={18}, description={robustness constant 
for stream \gls{stream} at port \gls{port}} }

\newglossaryentry{r_s2pw}
{ type=internal, name={\ensuremath{\Gamma}}, sort={18}, description={robustness constant 
for stream \gls{stream} at port \gls{port}} }

\newglossaryentry{r_sp2}
{ type=internal, name={\ensuremath{\Gamma}}, sort={18}, description={robustness constant 
for stream \gls{stream} at port \gls{port}} }

\newglossaryentry{ft_1p}
{ type=internal, name={\ensuremath{ft_s(1,\gls{port})}}, sort={18}, description={xx}}
\newglossaryentry{fd_1p}
{ type=internal, name={\ensuremath{fd_s(1,\gls{port})}}, sort={18}, description={xx}}

\newglossaryentry{ft_kp2}
{ type=internal, name={\ensuremath{ft_s(k,\gls{port_2})}}, sort={18}, description={xx}}
\newglossaryentry{fd_kp2}
{ type=internal, name={\ensuremath{fd_s(k,\gls{port_2})}}, sort={18}, description={xx}}

\newglossaryentry{func_tsp_2}
{ type=internal, name={\ensuremath{t(\gls{stream_2},\gls{port})}}, sort={16},
description={Total processing time required for \gls{stream_2} at port
\gls{port} , where other stream should not be assigned.} }

\newglossaryentry{func_tsp_2w}
{ type=internal, name={\ensuremath{t(\gls{stream_2},\gls{port_2})}}, sort={16},
description={Total processing time required for \gls{stream_2} at port
\gls{port} , where other stream should not be assigned.} }

\begin{document}

\title{A Robust Scheduling of Cyclic Traffic for Integrated Wired and Wireless Time-Sensitive Networks\thanks{This work was partly funded by the Bavarian State Government through the HighTech Agenda (HTA).}}

\makeatletter
\newcommand{\linebreakand}{%
  \end{@IEEEauthorhalign}
  \hfill\mbox{}\par
  \mbox{}\hfill\begin{@IEEEauthorhalign}
}
\author{
    \IEEEauthorblockN{
        Özgür Ozan Kaynak\IEEEauthorrefmark{1},
        Andreas Kassler\IEEEauthorrefmark{1},
        Andreas Fischer\IEEEauthorrefmark{1},
        Ognjen Dobrijevic\IEEEauthorrefmark{2},
        Fabio D'Andreagiovanni\IEEEauthorrefmark{3}
    }
    \IEEEauthorblockA{\IEEEauthorrefmark{1}Faculty of Computer Science, \textit{Deggendorf Institute of Technology (THD)}, Deggendorf, Germany\\
    Email: \{oezguer.kaynak, andreas.kassler, andreas.fischer\}@th-deg.de}
    \IEEEauthorblockA{\IEEEauthorrefmark{2}Department of Automation Technology, \textit{ABB Corporate Research}, Vasteras, Sweden \\
    Email: ognjen.dobrijevic@se.abb.com}
    \IEEEauthorblockA{\IEEEauthorrefmark{3}Department of Sciences and Method for Engineering, \textit{University of Modena and Reggio Emilia}, Reggio Emilia, Italy \\
    Email: fabio.dandreagiovanni@unimore.it}
}

\maketitle

\begin{abstract}
Time-Sensitive Networking (TSN) is a toolbox of technologies that enable deterministic communication over Ethernet. A key area has been TSN's time-aware traffic shaping (TAS), which supports stringent end-to-end latency and reliability requirements. Configuration of TAS requires the computation of a network-wide traffic schedule, which is particularly challenging with integrated wireless networks (e.g., 5G, Wi-Fi) due to the stochastic nature of wireless links. This paper introduces a novel method for configuring TAS, focusing on cyclic traffic patterns and jitter of wireless links. We formulate a linear program that computes a network-wide time-aware schedule, robust to wireless performance uncertainties. The given method enables robust scheduling of multiple TSN frames per transmission window using a tunable robustness parameter (\gls{r_sp}). To reduce computational complexity, we also propose a sequential batch-scheduling heuristic that runs in polynomial time. Our approach is evaluated by using different network topologies and wireless link characteristics, demonstrating that the heuristic can schedule 90\% of 6500 requested TSN streams in a large topology.
\end{abstract}

\section{Introduction}

Time-Sensitive Networking (TSN) initially started as a set of mechanisms and protocol extensions for Ethernet networks, developed by the IEEE 802.1 TSN Task Group~\cite{tsntaskgroup}. The applicability of TSN toolbox is being extended to 5G and Wi-Fi, but with a common goal: realizing deterministic communication over integrated wired--wireless networks. TSN supports critical applications in manufacturing, aviation, and the automotive industry, which demand communication timeliness and reliability. Data is typically sent via TSN bridges/switches in streams of layer-2 frames, from talker to listener end-stations. Cyclic data, such as industrial control traffic, is sent at fixed intervals (periods). For that purpose, IEEE 802.1AS time synchronization may be used to establish a common notion of time among all communicating TSN entities. A time-aware shaper (TAS), also known as IEEE 802.1Qbv, may then be employed to schedule network traffic in TSN talkers and bridges to meet the performance requirements of the respective streams. 

A main use case for TSN is to facilitate the adoption of Industry 4.0 principles by enabling a flexible network infrastructure—one capable of accommodating the diverse requirements of industrial applications as well as advances in robotics, machine learning, and edge computing. Integrating wireless access into wired TSN networks is critical for future smart factories, which increasingly rely on mobile industrial devices. Figure~\ref{fig1} illustrates a motivational scenario involving cooperative task execution by autonomous mobile robots and an unmanned aerial vehicle equipped with a video camera. A TSN-based infrastructure is employed to ensure timely data delivery for closed-loop control of the cooperating robots. To achieve high communication reliability, ground robots can establish connections to the TSN backbone via 5G and Wi-Fi.

\begin{figure}[hbtp]
\centerline{\includegraphics[width=\columnwidth]{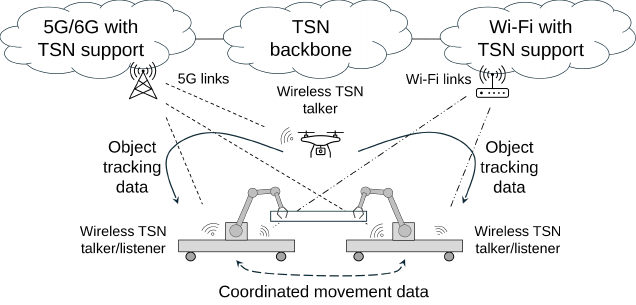}}
\caption{TSN over 5G and Wi-Fi for cooperative industrial automation}
\label{fig1}
\end{figure}

Wireless links are inherently more susceptible to resource contention and interference than wired links, posing greater challenges for communication determinism. Interference can cause frame or acknowledgment losses, triggering layer-2 retransmissions and increasing delay variability. These variable delays can cascade through the network, potentially leading to missed delivery deadlines and impacting other streams traversing shared TSN bridges~\cite{zanbouri2023comprehensive}. Standardization efforts have introduced features such as 5G Ultra-Reliable Low-Latency Communications (URLLC) to support deterministic and time-sensitive applications over wireless links~\cite{3gppURLLC}. Even with these advances, wireless segments still introduce uncertainty that complicates TAS configuration. Since a precise scheduling of gate opening and closing times, per egress port and queue, is required, a missed transmission window forces TSN frames to wait for the next cycle, amplifying jitter. While accounting for worst-case wireless (re)transmission delays can mitigate such an effect, it may be overly conservative, reducing efficiency or rendering schedules infeasible. Robust optimization offers a principled way to handle variability by calculating solutions feasible under data uncertainty, in turn enabling a reliable operation despite deviations from nominal values.

Addressing the aforementioned challenges of integrating wireless access into wired TSN networks, this paper focuses on scheduling cyclic traffic, as it is a dominant type in many industries. Our approach treats the wireless network as a black box, without configuring its internal working parameters (e.g., signal modulation or radio resource allocation). Instead, we assume that the wireless network designer provides a priori known lower and upper latency bounds for each wireless link, which are then used for deriving the network-wide TAS schedule. The paper contains the following contributions:
\begin{itemize}
    \item a novel, robust integer linear problem (ILP) formulation is proposed, for configuring TAS parameters in the presence of wireless performance uncertainties (based on our previous, initial work~\cite{kaynak2024}). A tunable parameter aims to balance robustness and scheduling capacity;
    \item a new, sequential batch-scheduling ILP heuristic is proposed, to reduce the problem complexity to polynomial time and enhance scalability to solve large problem instances fast; and
    \item our approach is evaluated using different network topology sizes and wireless link characteristics. We show that the exact model can only solve small problem sizes, while the given heuristic successfully schedules 90\% of 6500 requested TSN streams in a large topology.
\end{itemize}
The remainder of the paper is structured as follows. Section~II summarizes the state-of-the-art. The system model is described in Section~III. A detailed problem formulation and the heuristic design are presented in Section~IV. Section~V presents evaluation results and main findings, while Section~VI concludes the paper and sketches future work.

\section{Related Work}
The problem of scheduling cyclic traffic for TSN has been extensively studied, particularly for wired networks. Early approaches often cast the cyclic scheduling problem as a satisfiability problem~\cite{Craciunas2016} or an ILP~\cite{Durr2016}. Dürr \textit{et al.}~\cite{Durr2016} propose a no-wait scheduling approach where packets are transmitted from switches immediately upon arrival at egress ports. Craciunas \textit{et al.}~\cite{Craciunas2016} explore several scheduling strategies, including stream (flow) isolation, where frames of different streams are temporally isolated, and frames of the same stream are transmitted back-to-back. These foundational works typically assume deterministic link characteristics.

The integration of wireless links into TSN is an emerging research area driven by industrial needs. Zanbouri \textit{et al.}~\cite{zanbouri2023comprehensive} provide a comprehensive overview of the challenges and considerations when dealing with wireless uncertainties in TSN. Recent works by Egger \textit{et al.} \cite{egger2025end} consider a design where two wired networks are connected through a 5G logical bridge, focusing on schedulability and providing formal guarantees. Ginthör \textit{et al.}~\cite{Ginthor2020} proposed a constraint programming model to configure wired schedules along with 5G resources jointly, and they extended their work by focusing on wireless signal fading \cite{Ginthor2021}. Sharma \textit{et al.}~\cite{Sharma2024} model a two-stage (first routing, then scheduling) mixed ILP, considering an end-to-end wired and wireless no-wait scheduling approach. Li \textit{et al.}~\cite{lizhong2024} proposed a joint scheduling of TSN with Wi-Fi.

Robust Optimization offers a framework for handling data uncertainty in scheduling problems. Bertsimas and Sim introduced the $\Gamma$-robustness model~\cite{bertsimas} that allows users to configure system robustness via an adjustable parameter that controls the so-called "price of robustness", namely, the reduction in value that an optimal solution to the problem must face to be protected against data uncertainty.
The tunability of $\Gamma$-robustness is valuable for real-world applications where uncertainties, like variable wireless delays, are prevalent. Besides the easiness of tunability, which is highly desirable from a computational point of view,  we adopt Robust Optimization due to two other major advantages that it offers compared to other methods for optimization under data uncertainty, such as Stochastic Optimization: 1) embedding data uncertainty in the optimization model by defining a robust counterpart that preserves the nature of the original problem (e.g., a linear model admits a linear robust counterpart); 2) allowing to preserve the complexity of the original problem (see \cite{bertsimas}). 

Compared to previous works, our paper presents a robust ILP-based scheduling approach for integrated wired and wireless TSN that explicitly accounts for wireless delay uncertainty using a tunable robustness parameter. Additionally, we propose a scalable sequential batch-scheduling ILP heuristic that enables practical scheduling for large industrial TSNs.

\section{System Modeling and Assumptions}

We consider a TSN integrating wired and wireless segments (e.g., 5G, Wi-Fi). Wireless systems are treated as black boxes: their internal configuration is not modeled, but lower and upper delay bounds for each wireless link are assumed to be known (see Sec.~\ref{bounds}). These bounds are used to derive TAS schedules and talker offsets for the wired TSN switches. Our approach absorbs wireless uncertainties by dimensioning buffers at the wireless bridges, controlled by a parameter.

\subsection{Network Model}
The network is modeled as a directed graph $\gls{networkgraph} = (\gls{set_of_nodes}, \gls{set_of_links})$, where nodes $n \in N$ are wired/wireless end stations or bridges, and links $l \in L$ (wired or wireless) map to egress ports $\gls{port} \in \gls{set_of_ports}$. Scheduling focuses on Gate Control Lists (GCLs) configured by the TAS at each egress port, which control when queues can transmit. The scheduling problem determines GCLs and stream transmission offsets to meet all latency and jitter constraints~\cite{Stüber2023}. Streams traverse the network via links mapped to ports; wireless stations connect through dedicated wireless bridges. All nodes are time-synchronized. For wireless bridges, per-station buffers hold frames until their scheduled transmission time, providing functionality similar to TAS (e.g., de-jitter at a 5G node~\cite{Dinand2023patent}).

\subsection{Data Stream Model}
A set of TSN streams $\gls{stream} \in \gls{set_of_streams}$ is requested, each following a cyclic schedule from a talker to a listener (wired or wireless). Each stream has a fixed period $\gls{period_s}$, maximum latency $\gls{max_latency}$, and maximum jitter $\gls{max_jitter}$. Streams traverse a subset of nodes and links, corresponding to a subset of ordered ports $\gls{func_Ps} \subseteq \gls{set_of_ports}$ for a path $\gls{path}$. The cyclic schedule allocates distinct time windows for each stream at each port along its path. We model a single egress port queue class per port; queue classes can be configured freely in post-processing.

\section{Problem Model and Solution Design}

\begin{table}[htbp]
    \centering 
    
    \caption{Modelling Notations}
    \begin{tabular}{r|m{5.5cm}}
    Sets and Indices & Description\\
    \hline
    $\gls{set_of_streams}$ & Set of all data streams $\gls{stream}$. \\
    $\gls{set_of_ports}$ & Set of all egress ports $\gls{port}$. \\
    $\gls{path}$ & Index for a specific path of a \gls{stream}. \\
    $\gls{repetition}$ & Index for a repetition instance of a \gls{stream}. \\
    $\gls{set_of_paths}$ & Set of candidate paths for \gls{stream}. \\
    $\gls{func_Rs}$ & Set of repetition instances for \gls{stream} within the \gls{lcm}. \\
    $\gls{func_Ps}$ & Sequence of egress ports for \gls{stream} along \gls{path}. \\
    $\gls{func_P_inv_p}$ & Set of (\gls{stream}, \gls{path}) pairs that traverse \gls{port}. \\
    \hline
    Input & Description\\
    \hline
    $\gls{period_s}$ & Cycle period of a data stream \gls{stream}. \\
    $\gls{lcm}$ & Least common multiple of all \gls{period_s}, a greater configuration cycle time that will repeat in all devices. \\
    $\gls{max_latency}$ & Maximum end-to-end latency for stream \gls{stream}. \\
    $\gls{max_jitter}$ & Maximum jitter for a stream \gls{stream}, talker offset variation. \\
    $\gls{r_sp}$ & Robustness level parameter, configures uncertainty delay budget, a value in $[0, 1]$. \\
    $\gls{ft_kp}$ & Minimum time for the $k^{th}$ frame of \gls{stream} at \gls{port}. \\
    $\gls{fd_kp}$ & Maximum delay deviation for the $k^{th}$ frame of \gls{stream} at \gls{port}. \\
    $\gls{func_tsp}$ & Lower bound time to schedule for stream \gls{stream} at port \gls{port}. \\
    $\gls{func_dsp}$ & Delay deviation for stream \gls{stream} at port \gls{port}. \\
    $\gls{func_Nsp}$ & Gives the next port \gls{port_2} for stream \gls{stream} on path \gls{path} after current port \gls{port}. \\
    $\gls{func_ns_sp}$ & Required time offset for stream \gls{stream} between port \gls{port} and the next port \gls{port_2}. \\
    $\gls{fixed_set}$ & Function returning a set of fixed (\gls{fixed_start}, \gls{fixed_end}) time blocks on a \gls{port}. \\
    $\gls{fixed_start}$ & Start time of a pre-scheduled, fixed block $f$. \\
    $\gls{fixed_end}$ & End time of a pre-scheduled, fixed block $f$. \\
    $\gls{bigm}$ & A sufficiently large number for big-M constraints. \\

    \hline
    Decision Variables & Description\\
    \hline
    $\gls{dv_xrusp}$ & Start time of the transmission window for the $\gls{repetition}^{th}$ instance of \gls{stream} on \gls{path} at \gls{port}. \\
    $\gls{dv_as}$ & Binary variable: 1 if \gls{stream} is scheduled, 0 otherwise. \\
    $\gls{dv_zs}$ & Binary variable: 1 if \gls{stream} is scheduled on \gls{path}, 0 otherwise. \\
    $\gls{dv_y}$ & Binary helper variable for ordering streams in temporal isolation constraints. \\
    \end{tabular}
    \label{tab:notations}   
\end{table}

\subsection{Objective}
Given a set of requested streams $\gls{stream} \in \gls{set_of_streams}$, each with period $\gls{period_s}$, the goal is to find a schedule for all ports $\gls{port} \in \gls{set_of_ports}$. The schedule defines the path for each stream (as a sequence of ports), the talker offsets, and the GCL time windows for bridges.

\subsection{Model Inputs and Scheduling Design}

Times are represented as integers corresponding to TSN clock ticks. For each stream $\gls{stream}$, we compute up to $k$ candidate paths using Yen's k-shortest path algorithm~\cite{yens}. Each path is a sequence of ports $\gls{func_Ps}$ to be scheduled. The GCL configuration cycle is set to the least common multiple (\gls{lcm}) of all stream periods $\gls{period_s}$, with $|\gls{func_Rs}| = \gls{lcm} / \gls{period_s}$ stream instances per cycle.

Let $m$ be the number of frames a talker transmits for a stream within one cycle. For each frame $k$ at port $\gls{port}$, the minimum transmission time is:
\begin{equation} \label{eq:ft_kp}
    \gls{ft_kp} = trans_{k,\gls{port}} + prop_{k,\gls{port}} + proc_{k,\gls{port}}
\end{equation}
For wireless links, \gls{ft_kp} is equal to the best-case scenario (minimum delay). Wireless links introduce delay uncertainty, modeled by $\gls{fd_kp}$ (delay deviation); for wired links, $\gls{fd_kp}=0$. The cumulative minimum transmission time $\gls{func_tsp}$ and cumulative delay deviation $\gls{func_dsp}$ for all frames of a stream at a port are given by (\ref{eq:cumulative_times}):
\begin{equation} \label{eq:cumulative_times}
    \gls{func_tsp} = \sum_{k=1}^{m} \gls{ft_kp} \qquad
    \gls{func_dsp} = \sum_{k=1}^{m} \gls{fd_kp}
\end{equation}
The robustness parameter $\gls{r_sp} \in [0, 1]$ controls how much of the delay deviation is protected in the schedule: $\gls{r_sp}=0$ is optimistic, $\gls{r_sp}=1$ is fully robust. The forwarding is sequential: for each stream, the transmission of frame $i$ at the next port ($\gls{func_Nsp}=\gls{port_2}$) starts only after frame $i$ has been fully received and processed at the previous port (\gls{port}), and after frame $i-1$ has finished transmission at \gls{port_2}. This ensures frames are forwarded in order and never overlap. The offset time required $\gls{func_ns_sp}$ between consecutive ports is calculated (\ref{eq:ns_sp}):
\begin{equation} \label{eq:A_sp}
  A_i(\gls{stream}, \gls{port}) = \sum_{k=1}^{i} \gls{ft_kp} + \min\{\gls{r_sp} \cdot \gls{func_dsp}, \sum_{k=1}^{i} \gls{fd_kp}\}
\end{equation}
\begin{equation} \label{eq:B_sp}
    B(\gls{stream}, \gls{port}) = \max_{i=2}^{m} \left\{ A_i(\gls{stream}, \gls{port}) - A_{i-1}(\gls{stream}, \gls{port_2}) \right\}
\end{equation}
\begin{equation} \label{eq:ns_sp}
\gls{func_ns_sp}= \max\{A_1(\gls{stream}, \gls{port}), B(\gls{stream}, \gls{port})\}
\end{equation}
Frames are buffered at wireless bridges until their scheduled transmission window, creating deterministic departure times for subsequent scheduling. The robustness parameter $\gls{r_sp}$ directly scales the uncertainty budget $\gls{func_dsp}$, determining the required scheduling offset $\gls{func_ns_sp}$ between ports. Increasing $\gls{r_sp}$ allocates larger time windows for wireless transmissions, improving reliability but reducing network capacity for other streams. This tunable approach balances protection against wireless delay variation and overall schedulability.

\subsection{Integer Linear Program (ILP)}
With decision variables (stream: \gls{stream}, path: \gls{path}, port: \gls{port}, repetition instance within \gls{lcm}: \gls{repetition}),  \gls{dv_xrusp} is the transmission window starting times at ports for streams, (\gls{dv_zs}, \gls{dv_as}) are for (path, stream) selection (activation) and \gls{dv_y} is a variable to help building temporal isolation constraint (\ref{eq:no_overlap_constraint}):
\begin{equation*}
    \gls{dv_xrusp} \in \{0,\mathbb{Z}^+\}, \quad \{\gls{dv_y}, \gls{dv_zs}, \gls{dv_as}\} \in \{0, 1\}
\end{equation*}
The objective (left of (\ref{eq:objective_and_one_path})) is to maximize the number of scheduled streams, subject to the path selection constraint (right of (\ref{eq:objective_and_one_path})), which ensures that exactly one path is activated for each scheduled stream. Constraints, such as temporal isolation, latency, and jitter bounds, are detailed in constraints (\ref{eq:next_port_constraint})--(\ref{eq:max_jitter_constraint}).
\begin{equation}
    \max \sum_{\gls{stream} \in \gls{set_of_streams}} \gls{dv_as}
    \qquad
    \text{s.t.} \quad
    \sum_{\gls{path} \in \gls{set_of_paths}} \gls{dv_zs} = \gls{dv_as} \quad \forall \gls{stream} \in \gls{set_of_streams}
    \label{eq:objective_and_one_path}
\end{equation}
\vspace{-0.5cm}
\begin{align}
    \intertext{Constraint (\ref{eq:next_port_constraint}) enforces, given port \gls{port} and next port is \gls{port_2} of a path \gls{path} of stream \gls{stream}, \gls{port_2} time window is scheduled exactly after \gls{func_ns_sp} (\ref{eq:ns_sp}) amount of time:}
    & \begin{aligned}\label{eq:next_port_constraint}
        & \gls{dv_xrusp} + \gls{func_ns_sp}\cdot\gls{dv_zs} = \gls{dv_xrusp2} \\ 
        & \gls{stream} \in \gls{set_of_streams}, \gls{path} \in \gls{set_of_paths},  \gls{repetition} \in \gls{func_Rs},  \gls{port} \in  \gls{func_Ps}, \\
        & \gls{port_2} = \gls{func_Nsp}, \gls{port} \ne \textit{last}(\gls{func_Ps})
      \end{aligned}
\intertext{Constraint (\ref{eq:no_overlap_constraint}) isolates streams temporally at ports, considering robustness \gls{r_sp} and wireless delay deviations \gls{func_dsp}.}
    & \begin{aligned}\label{eq:no_overlap_constraint}
        & \text{(a)}\quad \gls{dv_xrusp} + (\gls{func_tsp} + \gls{func_dsp} \cdot \gls{r_sp}) \cdot \gls{dv_zs} \\
        &- \gls{dv_xrusp_2}  \leq  \gls{bigm} \cdot \gls{dv_y} \\
        & \text{(b)}\quad \gls{dv_xrusp_2} + (\gls{func_tsp_2} + \gls{func_ds2p} \cdot \gls{r_s2p} ) \cdot \gls{dv_zs2}   \\
        &- \gls{dv_xrusp} \leq \gls{bigm} \cdot (1 - \gls{dv_y})\\
        & \gls{stream} \in \gls{set_of_streams}, \gls{path} \in \gls{set_of_paths},  \gls{repetition} \in \gls{func_Rs},   \gls{port} \in  \gls{func_Ps}, \\ 
        & (\gls{stream_2}, \gls{path_2}) \in \gls{func_P_inv_p},  \gls{repetition_2} \in \gls{func_Rs_2}, \gls{stream} < \gls{stream_2} 
      \end{aligned} 
  \intertext{Constraint (\ref{eq:cyclic_periodic_constraint}), if a path of a stream is scheduled (\gls{dv_zs} = 1), a time window allocated for each stream instance at the talker port \gls{port} = \textit{first(\gls{func_Ps})} every \gls{period_s}, ensuring cyclic instance creation.}
    & \begin{aligned}\label{eq:cyclic_periodic_constraint}
        & \gls{dv_xrusp} \geq (\gls{repetition} - 1) \cdot \gls{period_s} \cdot \gls{dv_zs}   \\
        &  \gls{stream} \in \gls{set_of_streams}, \gls{path} \in \gls{set_of_paths}, \gls{repetition} \in \gls{func_Rs}, \gls{port} = \textit{first(\gls{func_Ps})} 
      \end{aligned}
    \intertext{Constraint (\ref{eq:max_latency_constraint}), end-to-end latency of stream from time=0  of cycle period \gls{period_s} is within the allotted  \gls{max_latency}.}
    & \begin{aligned}\label{eq:max_latency_constraint}
        & \gls{dv_xrusp} + (\gls{func_tsp} + \gls{func_dsp} \cdot \gls{r_sp}) \cdot \gls{dv_zs} \leq \\
        & \gls{max_latency} + (\gls{repetition} - 1) \cdot \gls{period_s}  \\
        &  \gls{stream} \in \gls{set_of_streams},  \gls{path} \in \gls{set_of_paths},  \gls{repetition} \in \gls{func_Rs}, \gls{port} = \textit{last(\gls{func_Ps})}
    \end{aligned}
    \intertext{Constraint (\ref{eq:max_jitter_constraint}), the jitter of a stream occurring from different instances \gls{repetition} of the same stream is bounded by \gls{max_jitter}.}
    & \begin{aligned}\label{eq:max_jitter_constraint}
        & \text{(a)}\quad (\gls{dv_xrusp} - (\gls{repetition} - 1) \cdot \gls{period_s}) - (\gls{dv_xr2usp} - (\gls{repetition_2} - 1) \cdot \gls{period_s}) \\
        & + \gls{bigm} \cdot (1 - \gls{dv_zs}) \geq - \gls{max_jitter} \\
        &\text{(b)}\quad (\gls{dv_xrusp} - (\gls{repetition} - 1) \cdot \gls{period_s}) - (\gls{dv_xr2usp} - (\gls{repetition_2} - 1) \cdot \gls{period_s}) \\
        & - \gls{bigm} \cdot (1 - \gls{dv_zs}) \leq \gls{max_jitter} \\
        &  \gls{stream} \in \gls{set_of_streams},  \gls{path} \in \gls{set_of_paths},\gls{port} = \textit{first(\gls{func_Ps})}, \gls{repetition} \in \gls{func_Rs}, \gls{repetition_2} \in \gls{func_Rs}, \gls{repetition} < \gls{repetition_2}
      \end{aligned}
\end{align}
\subsection{How to obtain link latency bounds in practice?}
\label{bounds}
Accurate wireless link delay bounds are essential but challenging, as radio frequency conditions and interference vary dynamically. Bounds can be obtained via empirical measurements (e.g., synchronized timestamps~\cite{det6gdata, URLLC_experiment}), vendor service level agreements (e.g., private 5G), or analytical methods like network calculus, though the latter may be overly conservative. Simulation-based profiling is useful in the design phase. In practice, combining offline profiling and online monitoring enables adaptive bounds, which is crucial in dynamic environments.
\subsection{Sequential Batch Scheduling Heuristic}
To address scalability, we propose a sequential batch scheduling ILP heuristic. Streams are partitioned into disjoint batches, and the ILP is solved iteratively for each batch. After each batch, scheduled streams are fixed and excluded from further optimization, reducing problem size but limiting global optimality. Batch size controls the trade-off between solution quality and computational time: smaller batches are faster but more myopic, while larger batches approach the exact ILP but are slower. Streams are sorted by scheduling priority, which reflects network preferences for which streams should be scheduled first. Thus, higher-priority streams are scheduled when resources are most available. Within each batch, the ILP schedules all streams without further prioritization.

In each iteration, fixed streams (from previous batches) have immutable paths and time windows, creating unavailable time blocks on ports. Variable streams (current batch) must avoid these blocks. To reduce constraints, consecutive fixed blocks too small for a variable stream are merged (see Figure~\ref{fig:fixed_time_windows}).

\begin{figure}[htbp]
    \centering
    \begin{tikzpicture}[x=0.6cm, y=1.6cm]
        \draw[->] (0,0) -- (12,0) node[right] {Time};

        \draw[fill=blue!30] (2,0.0) rectangle (4,0.3);
        \node[above] at (3.05,0.0) {\small Fixed \gls{stream_2}};

        \draw[fill=blue!30] (4.5,0.0) rectangle (7,0.3);
        \node[above] at (5.85,0.0) {\small Fixed \gls{stream_3}};

        \draw[|-|] (2,0.4) -- (7,0.4);
        \node[above] at (2,0.4) {\small \gls{fixed_start}};
        \node[above] at (7,0.4) {\small \gls{fixed_end}};

        \draw[fill=green!40, opacity=0.7] (8,0.0) rectangle (11,0.3);
        \node[above] at (9.5,0.0) {\small Variable \gls{stream}};

        \draw[decorate,decoration={brace,amplitude=5pt,mirror}] (0,0) -- (2,0) node[midway,below=6pt] {\small Available gap};
        \draw[decorate,decoration={brace,amplitude=5pt,mirror}] (4,0) -- (4.5,0) node[midway,below=6pt] {\small Small gap};
        \draw[decorate,decoration={brace,amplitude=5pt,mirror}] (7,0) -- (12,0) node[midway,below=6pt] {\small Available gap};
    \end{tikzpicture}
    \caption{\gls{fixed_start} and \gls{fixed_end} times for a variable stream}
    \label{fig:fixed_time_windows}
\end{figure}
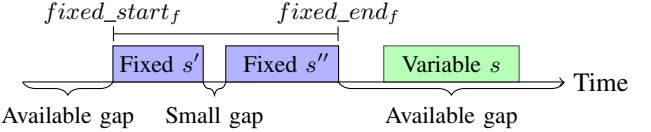

Let \gls{fixed_set} be a function that returns the set of merged fixed time blocks $(\gls{fixed_start}, \gls{fixed_end})$ for a variable stream \gls{stream} at port \gls{port}, representing unavailable temporal windows due to previously scheduled streams (see Figure~\ref{fig:fixed_time_windows}). Constraint~(\ref{eq:fixed_no_overlap_constraint}) ensures that the transmission window of a new variable stream does not overlap with any fixed block on the port; it must be scheduled entirely before or after each fixed block. Algorithm~\ref{alg:heuristic} outlines the sequential scheduling process.

\begin{align}
    & \begin{aligned}\label{eq:fixed_no_overlap_constraint}
    & \text{(a)} \quad \gls{dv_xrusp} + (\gls{func_tsp} + \gls{func_dsp} \cdot \gls{r_sp}) \cdot \gls{dv_zs} \leq \\ 
    & \gls{fixed_start} + \gls{bigm} \cdot \gls{dv_yf} \\
    & \text{(b)} \quad \gls{fixed_end} \leq \gls{dv_xrusp} + \gls{bigm} \cdot (1 - \gls{dv_yf}) \\
    &  \gls{stream} \in \gls{set_of_streams},  \gls{path} \in \gls{set_of_paths},  \gls{repetition} \in \gls{func_Rs},  \gls{port} \in \gls{func_Ps} \\ 
    & (\gls{fixed_start}, \gls{fixed_end}) \in \gls{fixed_set} \\
    \end{aligned}
\end{align}
\begin{algorithm}[htbp]
    \caption{Sequential Batch Scheduling Heuristic}
    \label{alg:heuristic}
    \KwIn{Set of streams $\mathcal{S}$, number of batches $B$}
    \KwOut{Fixed variables, \gls{dv_zs}, \gls{dv_as}, \gls{dv_xrusp}}
    Sort streams in $\mathcal{S}$ by priority into $\mathcal{S}_{sorted}$\;
    Partition $\mathcal{S}_{sorted}$ into $B$ disjoint batches: $\mathcal{B}_1, \ldots, \mathcal{B}_B$\;
    \For{$i \gets 1$ \KwTo $B$}{
        Set streams in $\mathcal{B}_i$ as variables, all previously scheduled streams as fixed\;
        Reset variables \gls{dv_y}, \gls{dv_yf}\;
        Reset and re-add constraints (\ref{eq:no_overlap_constraint}) and (\ref{eq:fixed_no_overlap_constraint}) for $\mathcal{B}_i$ \;
        Solve the ILP for the current batch\;
        Fix variables \gls{dv_as}, \gls{dv_zs}, \gls{dv_xrusp} for scheduled streams\;
    }
\end{algorithm}

The computational complexity grows with the number of streams sharing a port, especially for variable streams. Fixed streams add smaller complexity, and merging fixed blocks reduces constraints as scheduling progresses. The heuristic scales polynomially by iterating over fixed streams to build sequential ILP models, with batch size limiting the complexity of ILPs. This enables efficient scheduling for large TSNs.
\section{Evaluation}
Our evaluation aims to answer the following questions:
\begin{enumerate}
    \item How does the heuristic compare against the optimal? What is the impact of different link characteristics, network topologies, streams, and batch sizes? (cf.~\ref{eval:comp})
    \item What is the trade-off between robustness, wireless link latency deviations, and the number of scheduled streams? (cf.~\ref{eval:robust})
\end{enumerate}
\subsection{Experimental Setup}
\subsubsection{Evaluation Platform}
Experiments ran on a server with an Intel Xeon Gold 6326 CPU (32 cores) and 250 GB RAM. We used Gurobi Optimizer 12.0.1 with a 2-hour time limit for medium and small, and 4-hour for large per experiment to solve ILPs. For the exact ILP, this limit applies to the single run; for the batch heuristic, it is distributed equally across all batches. If the time limit is reached, the best feasible solution is used. The code and experiments we share for reproducibility.\footnote{\url{https://mygit.th-deg.de/inets/rctsn}}

\subsubsection{Network and Link Characteristics}
Wired links use 100 Mbps, with $propagation\_delay$ and $processing\_delay$ set to 10$\mu$s (see (1)). For wireless, we use two datasets: URLLC~\cite{URLLC_experiment} (latencies for 32B and 1420B packets, extrapolated for others; see Figure~\ref{fig:urllc_histogams}) and Det6G~\cite{det6gdata} (5G, long-tail delay, see Figure~\ref{fig:det6g_histograms}). Histograms determine \gls{ft_kp} (minimum observed latency) and \gls{fd_kp} (delay deviation, i.e., the difference between the upper and lower bounds of the link latency).

\begin{figure}[!t]
    \centering
    \includegraphics[width=\linewidth]{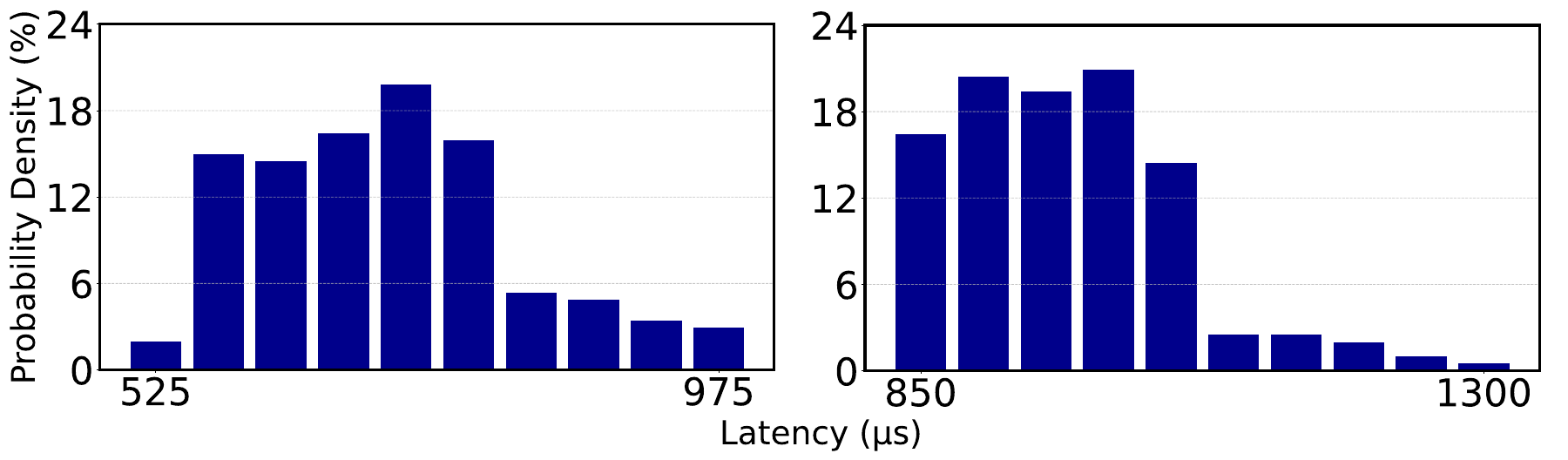}
    \setlength{\belowcaptionskip}{-8pt}
    \vspace{-0.25cm}
    \caption{Wireless link delay histograms for different packet sizes (URLLC scenario, left: 32B, right: 1420B)~\cite{URLLC_experiment}}
    \label{fig:urllc_histogams}
\end{figure}

\begin{figure}[!t]
    \centering
    \includegraphics[width=\linewidth]{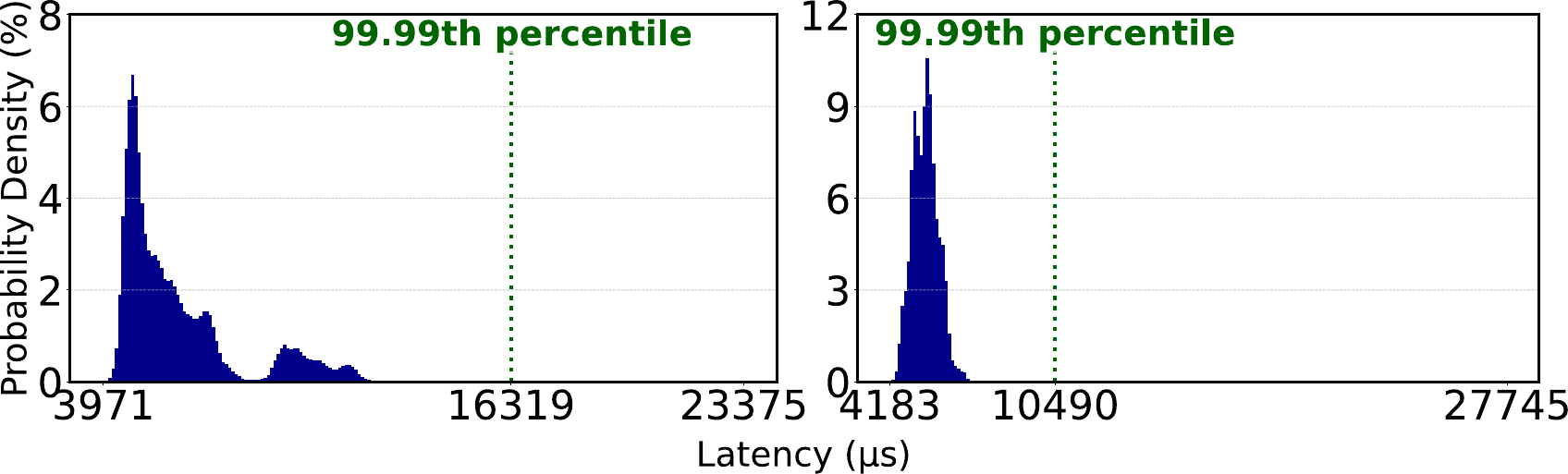}
    \setlength{\belowcaptionskip}{-8pt}
    \vspace{-0.25cm}
    \caption{Wireless link delay histograms for 100B packets (Det6G data, left: downlink, right: uplink)~\cite{det6gdata}}
    \label{fig:det6g_histograms}
\end{figure}

\subsubsection{Traffic Scenarios}
TSN streams are generated by randomly selecting properties from Table~\ref{tab:stream_params}. Each stream has \gls{max_latency} and \gls{max_jitter} that are proportional random multiples of \gls{period_s}. Streams have priority levels 1--3 (equal distribution); priority 3 is highest. Det6G scenarios designed with longer \gls{period_s} due to higher latencies.

\begin{table}[!t]
    \centering
    \caption{Traffic and Stream Parameters}
    \begin{tabular}{l|p{5cm}}
        \gls{period_s} (URLLC) & $\{500, 1000, 2000, 4000, 8000\}$~$\mu$s \\
        \hline
        $packet\_size$ (URLLC) & $\{32, 64, 128, 256, 512, 1024, 1420\}$~Bytes \\
        \hline
        \gls{period_s} (Det6G) & $\{20000, 40000\}$~$\mu$s \\
        \hline
        $packet\_size$ (Det6G) & $\{100\}$~Bytes \\
        \hline
        \gls{max_latency} & \{0.5, 0.6, 0.7, 0.8, 1.0\} times \gls{period_s} \\
        \hline
        \gls{max_jitter} & \{0.1, 0.2, 0.3, 0.4, 0.5\} times \gls{period_s} \\
        \hline
        $priority$ & \{1, 2, 3\}, highest is 3, distributed equally \\
    \end{tabular}
    \label{tab:stream_params}
\end{table}

\subsubsection{Experimental Design}
We test three topologies: small (5 wired, 5 wireless bridges, ring), medium (50, 50), and large (220, 220), with random sub-topologies. For computational experiments, we use URLLC, varying batch size and stream count, fixing \gls{r_sp}=1. For the robustness evaluation, we used the small topology with both the URLLC and Det6G scenarios, employing only the exact ILP to ensure a fair comparison.

For each stream, up to $k=3$ shortest paths are precalculated using Yen's algorithm~\cite{yens}. Experimental stream sets are generated using the heuristic with small batches, then shuffled (except for priority, which is uniform). This biases the experimental sets toward less stringent requirements, but some tight-deadline streams remain. In theory, the exact model can schedule all streams given unlimited resources.

\subsubsection{Key Performance Indicators}
Streams scheduled (\%) is the ratio of scheduled to requested streams; priority 3 scheduled (\%) is the ratio for priority 3 streams. Port utilization (\%) is the average fraction of time ports are scheduled to transmit in \gls{lcm}. Utilization rises with more streams and longer paths. Success probability (\%) is the cumulative delay distribution up to the chosen robustness level \gls{r_sp} and upper bound \gls{func_dsp}, representing the likelihood that wireless transmissions finish within the allocated window; for multiple wireless links, probabilities are multiplied.
\subsection{Computational Experiments}
\label{eval:comp}
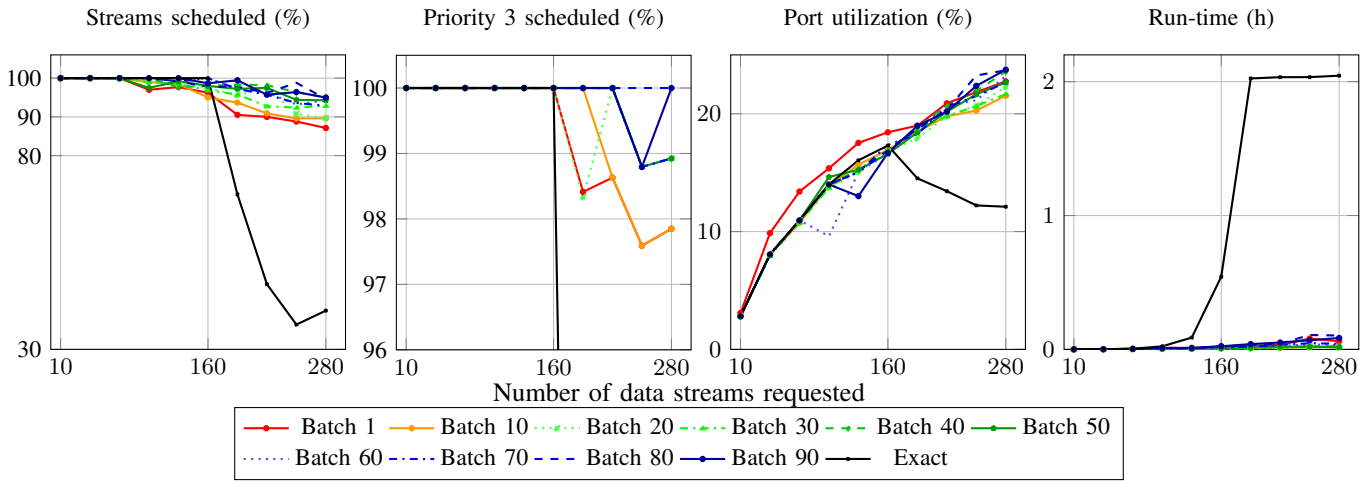
\begin{figure*}[!t]
    \centering
    \begin{minipage}{0.24\textwidth}
    \centering
    \begin{tikzpicture}
        \begin{axis}[
            width=5.5cm, height=5.5cm,
            title = {\small Streams scheduled (\%)},
            legend to name=smallLegend,
            legend columns=6,
            legend style={font=\small},
            grid=major,
            xmin=0, xmax=300,
            ymin=0.3, ymax=1.06,
            xtick={10, 160, 280},
            ytick={0.30, 0.80, 0.90,  1.0},
            yticklabels={30, 80, 90, 100},
            yticklabel style={/pgf/number format/fixed, /pgf/number format/precision=1},
            tick label style={font=\small},
            label style={font=\small},
        ]
        \addplot[mark=oplus*, mark size=0.8pt, thick, color=rangeC4] coordinates { (10,1.0) (40,1.0) (70,1.0) (100,0.97) (130,0.9769230769230769) (160,0.9625) (190,0.9052631578947369) (220,0.9) (250,0.888) (280,0.8714285714285714) }; \addlegendentry{Batch 1}
        \addplot[mark=*, mark size=0.8pt, thick, color=rangeB4] coordinates { (10,1.0) (40,1.0) (70,1.0) (100,0.99) (130,0.9846153846153847) (160,0.95) (190,0.9368421052631579) (220,0.909090909090909) (250,0.896) (280,0.8964285714285715) }; \addlegendentry{Batch 10}
        \addplot[mark=square*, mark size=0.8pt, thick, dotted, color=rangeD3] coordinates { (10,1.0) (40,1.0) (70,1.0) (100,0.99) (130,0.9833333333333333) (160,0.975) (190,0.9555555555555556) (220,0.975) (250,0.908) (280,0.8964285714285715) }; \addlegendentry{Batch 20}
        \addplot[mark=triangle*, mark size=0.8pt, thick, dashdotted, color=rangeD4] coordinates { (10,1.0) (40,1.0) (70,1.0) (100,0.99) (130,0.9846153846153847) (160,0.9687500000000001) (190,0.9578947368421052) (220,0.9272727272727272) (250,0.924) (280,0.9285714285714286) }; \addlegendentry{Batch 30}
        \addplot[mark=diamond*, mark size=0.8pt, thick, dashed, color=rangeD5] coordinates { (10,1.0) (40,1.0) (70,1.0) (100,1.0) (130,0.9769230769230769) (160,0.98) (190,0.9833333333333333) (220,0.9818181818181818) (250,0.964) (280,0.95) }; \addlegendentry{Batch 40}
        \addplot[mark=pentagon*, mark size=0.8pt, thick, color=rangeD6] coordinates { (10,1.0) (40,1.0) (70,1.0) (100,0.975) (130,0.9916666666666667) (160,0.98) (190,0.9722222222222222) (220,0.975) (250,0.944) (280,0.9428571428571428) }; \addlegendentry{Batch 50}
        \addplot[mark=star*, mark size=0.8pt, thick, dotted, color=rangeE3] coordinates { (10,1.0) (40,1.0) (70,1.0) (100,1.0) (130,0.9916666666666667) (160,1.0) (190,0.9722222222222222) (220,0.9633802816901409) (250,0.932) (280,0.9428571428571428) }; \addlegendentry{Batch 60}
        \addplot[mark=10-pointed star*, mark size=0.8pt, thick, dashdotted, color=rangeE4] coordinates { (10,1.0) (40,1.0) (70,1.0) (100,1.0) (130,0.9916666666666667) (160,1.0) (190,0.9722222222222222) (220,0.9571428571428572) (250,0.936) (280,0.9285714285714286) }; \addlegendentry{Batch 70}
        \addplot[mark=Mercedes star*, mark size=0.8pt, thick, dashed, color=rangeE5] coordinates { (10,1.0) (40,1.0) (70,1.0) (100,1.0) (130,0.9916666666666667) (160,0.9777777777777777) (190,0.9777777777777777) (220,0.9619047619047619) (250,0.988) (280,0.9464285714285714) }; \addlegendentry{Batch 80}
        \addplot[mark=otimes*, mark size=0.8pt, thick, color=rangeE6] coordinates { (10,1.0) (40,1.0) (70,1.0) (100,1.0) (130,1.0) (160,0.9866666666666667) (190,0.9944444444444445) (220,0.9571428571428572) (250,0.964) (280,0.95) }; \addlegendentry{Batch 90}
        \addplot[mark=x, mark size=0.8pt, thick, color=exact] coordinates { (10,1.0) (40,1.0) (70,1.0) (100,1.0) (130,1.0) (160,1.0) (190,0.7) (220,0.4681818181818182) (250,0.364) (280,0.4) }; \addlegendentry{Exact}
        \end{axis}
    \end{tikzpicture}
    \end{minipage}\hfill
    \begin{minipage}{0.24\textwidth}
        \centering
        \begin{tikzpicture}
            \begin{axis}[
                width=5.5cm, height=5.5cm,
                title={\small Priority 3 scheduled (\%)},
                grid=major, xmin=0, xmax=300, ymin=0.96, ymax=1.005,
                xtick={10, 160, 280}, 
                ytick={0.96, 0.97, 0.98, 0.99, 1.0},
                yticklabels={96, 97, 98, 99, 100},
                yticklabel style={/pgf/number format/fixed, /pgf/number format/precision=1},
                tick label style={font=\small}, label style={font=\small},
            ]
            \addplot[mark=oplus*, mark size=0.8pt, thick, color=rangeC4] coordinates { (10,1.0) (40,1.0) (70,1.0) (100,1.0) (130,1.0) (160,1.0) (190,0.9841269841269841) (220,0.9863013698630137) (250,0.9759036144578313) (280,0.9784946236559140) };
            \addplot[mark=*, mark size=0.8pt, thick, color=rangeB4] coordinates { (10,1.0) (40,1.0) (70,1.0) (100,1.0) (130,1.0) (160,1.0) (190,1.0) (220,0.9863013698630137) (250,0.9759036144578313) (280,0.9784946236559140) };
            \addplot[mark=square*, mark size=0.8pt, thick, dotted, color=rangeD3] coordinates { (10,1.0) (40,1.0) (70,1.0) (100,1.0) (130,1.0) (160,1.0) (190,0.9833333333333333) (220,1.0) (250,0.9879518072289157) (280,0.989247311827957) };
            \addplot[mark=triangle*, mark size=0.8pt, thick, dashdotted, color=rangeD4] coordinates { (10,1.0) (40,1.0) (70,1.0) (100,1.0) (130,1.0) (160,1.0) (190,1.0) (220,1.0) (250,0.9879518072289157) (280,0.9892473118279570) };
            \addplot[mark=diamond*, mark size=0.8pt, thick, dashed, color=rangeD5] coordinates { (10,1.0) (40,1.0) (70,1.0) (100,1.0) (130,1.0) (160,1.0) (190,1.0) (220,1.0) (250,0.9879518072289157) (280,0.989247311827957) };
            \addplot[mark=pentagon*, mark size=0.8pt, thick, color=rangeD6] coordinates { (10,1.0) (40,1.0) (70,1.0) (100,1.0) (130,1.0) (160,1.0) (190,1.0) (220,1.0) (250,0.9879518072289157) (280,0.989247311827957) };
            \addplot[mark=star*, mark size=0.8pt, thick, dotted, color=rangeE3] coordinates { (10,1.0) (40,1.0) (70,1.0) (100,1.0) (130,1.0) (160,1.0) (190,1.0) (220,1.0) (250,0.9879518072289157) (280,0.989247311827957) };
            \addplot[mark=10-pointed star*, mark size=0.8pt, thick, dashdotted, color=rangeE4] coordinates { (10,1.0) (40,1.0) (70,1.0) (100,1.0) (130,1.0) (160,1.0) (190,1.0) (220,1.0) (250,0.9879518072289157) (280,0.989247311827957) };
            \addplot[mark=Mercedes star*, mark size=0.8pt, thick, dashed, color=rangeE5] coordinates { (10,1.0) (40,1.0) (70,1.0) (100,1.0) (130,1.0) (160,1.0) (190,1.0) (220,1.0) (250,1.0) (280,1.0) };
            \addplot[mark=otimes*, mark size=0.8pt, thick, color=rangeE6] coordinates { (10,1.0) (40,1.0) (70,1.0) (100,1.0) (130,1.0) (160,1.0) (190,1.0) (220,1.0) (250,0.9879518072289157) (280,1.0) };
            \addplot[mark=x, mark size=0.8pt, thick, color=exact] coordinates { (10,1.0) (40,1.0) (70,1.0) (100,1.0) (130,1.0) (160,1.0) (190,0.7460317460317460) (220,0.3972602739726027) (250,0.3373493975903614) (280,0.3978494623655914) };
            \end{axis}
        \end{tikzpicture}
    \end{minipage}\hfill
    \begin{minipage}{0.24\textwidth}
        \centering
        \begin{tikzpicture}
            \begin{axis}[
                width=5.5cm, height=5.5cm,
                title={\small Port utilization (\%)},
                grid=major, xmin=0, xmax=300, ymin=0, ymax=25, 
                xtick={10, 160, 280},
                ytick={0, 10, 20}, 
                tick label style={font=\small}, label style={font=\small},
            ]
            \addplot[mark=oplus*, mark size=0.8pt, thick, color=rangeC4] coordinates { (10,3.09) (40,9.87) (70,13.4) (100,15.37) (130,17.53) (160,18.43) (190,18.99) (220,20.89) (250,21.84) (280,22.75) };
            \addplot[mark=*, mark size=0.8pt, thick, color=rangeB4] coordinates { (10,2.81) (40,8.03) (70,10.74) (100,13.87) (130,15.68) (160,16.93) (190,18.49) (220,19.79) (250,20.27) (280,21.53) };
            \addplot[mark=square*, mark size=0.8pt, thick, dotted, color=rangeD3] coordinates { (10,2.81) (40,8.06) (70,10.95) (100,13.73) (130,15.04) (160,16.84) (190,17.91) (220,19.8) (250,20.71) (280,22.29) };
            \addplot[mark=triangle*, mark size=0.8pt, thick, dashdotted, color=rangeD4] coordinates { (10,2.81) (40,7.94) (70,10.76) (100,13.73) (130,15.32) (160,16.67) (190,18.66) (220,19.8) (250,20.65) (280,21.67) };
            \addplot[mark=diamond*, mark size=0.8pt, thick, dashed, color=rangeD5] coordinates { (10,2.81) (40,8.06) (70,10.95) (100,14.0) (130,15.25) (160,16.92) (190,18.85) (220,20.68) (250,21.94) (280,23.58) };
            \addplot[mark=pentagon*, mark size=0.8pt, thick, color=rangeD6] coordinates { (10,2.81) (40,8.06) (70,10.95) (100,14.63) (130,15.25) (160,16.6) (190,18.4) (220,20.18) (250,21.65) (280,22.67) };
            \addplot[mark=star*, mark size=0.8pt, thick, dotted, color=rangeE3] coordinates { (10,2.81) (40,8.06) (70,10.95) (100,9.61) (130,15.05) (160,16.64) (190,18.73) (220,20.71) (250,21.14) (280,23.21) };
            \addplot[mark=10-pointed star*, mark size=0.8pt, thick, dashdotted, color=rangeE4] coordinates { (10,2.81) (40,8.06) (70,10.95) (100,14.0) (130,15.05) (160,17.35) (190,18.28) (220,20.55) (250,21.51) (280,22.66) };
            \addplot[mark=Mercedes star*, mark size=0.8pt, thick, dashed, color=rangeE5] coordinates { (10,2.81) (40,8.06) (70,10.95) (100,14.0) (130,15.04) (160,16.92) (190,18.71) (220,20.34) (250,23.25) (280,23.69) };
            \addplot[mark=otimes*, mark size=0.8pt, thick, color=rangeE6] coordinates { (10,2.81) (40,8.06) (70,10.95) (100,14.0) (130,13.01) (160,16.68) (190,18.97) (220,20.18) (250,22.37) (280,23.74) };
            \addplot[mark=x, mark size=0.8pt, thick, color=exact] coordinates { (10,2.81) (40,8.06) (70,10.95) (100,14.0) (130,16.05) (160,17.3) (190,14.52) (220,13.43) (250,12.22) (280,12.11) };
            \end{axis}
        \end{tikzpicture}
    \end{minipage}\hfill
    \begin{minipage}{0.24\textwidth}
        \centering
        \begin{tikzpicture}
            \begin{axis}[
                width=5.5cm, height=5.5cm,
                title={\small Run-time (h)},
                grid=major, xmin=0, xmax=300, ymin=0, ymax=2.2, 
                xtick={10, 160, 280},
                ytick={0, 1, 2}, 
                tick label style={font=\small}, label style={font=\small},
            ]
            \addplot[mark=oplus*, mark size=0.8pt, thick, color=rangeC4] coordinates { (10,0.000214) (40,0.001324) (70,0.005102) (100,0.011458) (130,0.010756) (160,0.022694) (190,0.026708) (220,0.033614) (250,0.080675) (280,0.060731) };
            \addplot[mark=*, mark size=0.8pt, thick, color=rangeB4] coordinates { (10,0.000180) (40,0.000500) (70,0.001456) (100,0.003319) (130,0.005200) (160,0.007472) (190,0.010750) (220,0.012092) (250,0.016792) (280,0.021153) };
            \addplot[mark=square*, mark size=0.8pt, thick, dotted, color=rangeD3] coordinates { (10,0.000296) (40,0.000498) (70,0.001272) (100,0.004992) (130,0.005583) (160,0.006257) (190,0.007082) (220,0.009926) (250,0.015748) (280,0.013915) };
            \addplot[mark=triangle*, mark size=0.8pt, thick, dashdotted, color=rangeD4] coordinates { (10,0.000275) (40,0.000560) (70,0.002036) (100,0.004356) (130,0.005611) (160,0.009794) (190,0.010372) (220,0.014922) (250,0.018653) (280,0.019903) };
            \addplot[mark=diamond*, mark size=0.8pt, thick, dashed, color=rangeD5] coordinates { (10,0.000277) (40,0.001294) (70,0.004578) (100,0.006462) (130,0.010862) (160,0.006845) (190,0.010286) (220,0.016672) (250,0.020886) (280,0.024018) };
            \addplot[mark=pentagon*, mark size=0.8pt, thick, color=rangeD6] coordinates { (10,0.000275) (40,0.001329) (70,0.001784) (100,0.005450) (130,0.005451) (160,0.009900) (190,0.014969) (220,0.018819) (250,0.021289) (280,0.015006) };
            \addplot[mark=star*, mark size=0.8pt, thick, dotted, color=rangeE3] coordinates { (10,0.000276) (40,0.001181) (70,0.001726) (100,0.008724) (130,0.007819) (160,0.009133) (190,0.023895) (220,0.025875) (250,0.035985) (280,0.033928) };
            \addplot[mark=10-pointed star*, mark size=0.8pt, thick, dashdotted, color=rangeE4] coordinates { (10,0.000108) (40,0.001155) (70,0.001728) (100,0.006617) (130,0.007819) (160,0.015619) (190,0.020539) (220,0.033103) (250,0.046064) (280,0.042289) };
            \addplot[mark=Mercedes star*, mark size=0.8pt, thick, dashed, color=rangeE5] coordinates { (10,0.000134) (40,0.001155) (70,0.002159) (100,0.010861) (130,0.010862) (160,0.023169) (190,0.031147) (220,0.042420) (250,0.107440) (280,0.103453) };
            \addplot[mark=otimes*, mark size=0.8pt, thick, color=rangeE6] coordinates { (10,0.000917) (40,0.001155) (70,0.004020) (100,0.010250) (130,0.010250) (160,0.023169) (190,0.039470) (220,0.051372) (250,0.068285) (280,0.085124) };
            \addplot[mark=x, mark size=0.8pt, thick, color=exact] coordinates { (10,0.000075) (40,0.001155) (70,0.004578) (100,0.023119) (130,0.088400) (160,0.543142) (190,2.024028) (220,2.033386) (250,2.033944) (280,2.044664) };
            \end{axis}
        \end{tikzpicture}
    \end{minipage}\hfill
    \centerline{Number of data streams requested}
    \centerline{\ref{smallLegend}}
    \caption{Small topology computational experiments}
    \label{fig:computational_results_small}
\end{figure*}
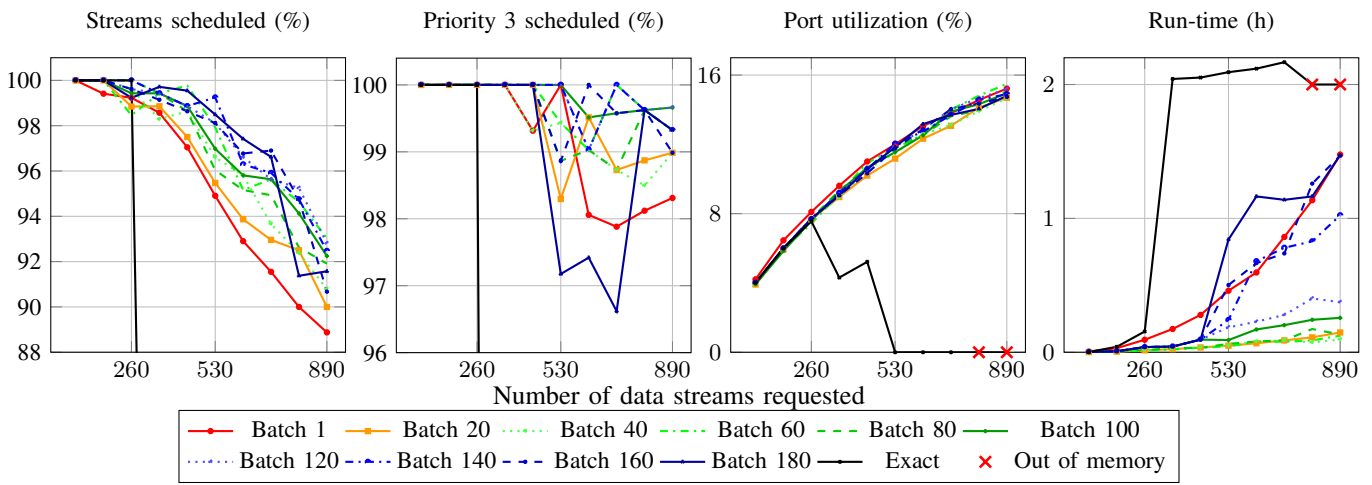
\begin{figure*}[!t]
    \centering
    \begin{minipage}{0.24\textwidth}
    \centering
    \begin{tikzpicture}
        \begin{axis}[
            width=5.5cm, height=5.5cm,
            title = {\small Streams scheduled (\%)},
            legend to name=mediumLegend,
            legend columns=6,
            legend style={font=\small},
            grid=major, xmin=0, xmax=950, ymin=0.88, ymax=1.01,
            xtick={260, 530, 890}, 
            ytick={0.88,0.90,0.92,0.94,0.96,0.98, 1.0},
            yticklabels={88,90,92,94,96,98, 100},
            yticklabel style={/pgf/number format/fixed, /pgf/number format/precision=1},
            tick label style={font=\small}, label style={font=\small},
        ]
        \addplot[mark=oplus*, mark size=0.8pt, thick, color=rangeC4] coordinates { (80,1.0) (170,0.9941176470588236) (260,0.9923076923076923) (350,0.9857142857142858) (440,0.9704545454545455) (530,0.9490566037735849) (620,0.9290322580645162) (710,0.9154929577464789) (800,0.9) (890,0.8887640449438202) }; \addlegendentry{Batch 1}
        \addplot[mark=square*, mark size=0.8pt, thick, color=rangeB4] coordinates { (80,1.0) (170,1.0) (260,0.9884615384615385) (350,0.9885714285714285) (440,0.975) (530,0.9547169811320755) (620,0.9387096774193548) (710,0.9295774647887324) (800,0.925) (890,0.9) }; \addlegendentry{Batch 20}
        \addplot[mark=diamond*, mark size=0.8pt, thick, dotted, color=rangeD3] coordinates { (80,1.0) (170,1.0) (260,0.9961538461538462) (350,0.9828571428571429) (440,0.9863636363636363) (530,0.9660377358490566) (620,0.9580645161290322) (710,0.9366197183098591) (800,0.92375) (890,0.9078651685393259) }; \addlegendentry{Batch 40}
        \addplot[mark=star*, mark size=0.8pt, thick, dashdotted, color=rangeD4] coordinates { (80,1.0) (170,1.0) (260,0.9846153846153847) (350,0.9942857142857143) (440,0.9977272727272727) (530,0.9792452830188679) (620,0.9516129032258065) (710,0.956338028169014) (800,0.94625) (890,0.9303370786516854) }; \addlegendentry{Batch 60}
        \addplot[mark=Mercedes star*, mark size=0.8pt, thick, dashed, color=rangeD5] coordinates { (80,1.0) (170,1.0) (260,1.0) (350,0.9942857142857143) (440,0.9863636363636363) (530,0.960377358490566) (620,0.9516129032258065) (710,0.9492957746478873) (800,0.92625) (890,0.9191011235955057) }; \addlegendentry{Batch 80}
        \addplot[mark=+, mark size=0.8pt, thick, color=rangeD6] coordinates { (80,1.0) (170,1.0) (260,0.9942857142857143) (350,0.9942857142857143) (440,0.98863636363636) (530,0.969811320754717) (620,0.9580645161290322) (710,0.956338028169014) (800,0.94125) (890,0.9224719101123595) }; \addlegendentry{Batch 100}
        \addplot[mark=triangle, mark size=0.8pt, thick, dotted, color=rangeE3] coordinates { (80,1.0) (170,1.0) (260,0.9961538461538462) (350,0.9942857142857143) (440,0.9886363636363636) (530,0.9811320754716981) (620,0.9661290322580646) (710,0.956338028169014) (800,0.9525) (890,0.9280898876404494) }; \addlegendentry{Batch 120}
        \addplot[mark=o, mark size=0.8pt, thick, dashdotted, color=rangeE4] coordinates { (80,1.0) (170,1.0) (260,1.0) (350,0.9942857142857143) (440,0.9886363636363636) (530,0.9924528301886792) (620,0.9629032258064516) (710,0.9591549295774648) (800,0.9475) (890,0.9247191011235955) }; \addlegendentry{Batch 140}
        \addplot[mark=star, mark size=0.8pt, thick, dashed, color=rangeE5] coordinates { (80,1.0) (170,1.0) (260,0.9961538461538462) (350,0.9914285714285714) (440,0.9863636363636363) (530,0.9811320754716981) (620,0.967741935483871) (710,0.9690140845070423) (800,0.9475) (890,0.9067415730337078) }; \addlegendentry{Batch 160}
        \addplot[mark=Mercedes star, mark size=0.8pt, thick, color=rangeE6] coordinates { (80,1.0) (170,1.0) (260,0.9923076923076923) (350,0.9971428571428571) (440,0.9954545454545455) (530,0.9849056603773585) (620,0.9741935483870968) (710,0.9661971830985916) (800,0.91375) (890,0.9157303370786517) }; \addlegendentry{Batch 180}
        \addplot[mark=x, mark size=0.8pt, thick, color=exact] coordinates { (80,1.0) (170,1.0) (260,1.0) (350,0.3514285714285714) (440,0.0022727272727272726) (530,0.0) (620,0.0) (710,0.0) }; \addlegendentry{Exact}
        \addplot[only marks, mark=x, mark options={scale=1.5, color=red, line width=1pt}] coordinates { (800,0.0) (890,0.0) }; \addlegendentry{Out of memory}
        \end{axis}
    \end{tikzpicture}
    \end{minipage}\hfill
    \begin{minipage}{0.24\textwidth}
        \centering
        \begin{tikzpicture}
            \begin{axis}[
                width=5.5cm, height=5.5cm,
                title = {\small Priority 3 scheduled (\%)},
                grid=major, xmin=0, xmax=950, ymin=0.96, ymax=1.004,
                xtick={260, 530, 890}, 
                ytick={0.96,0.97,0.98,0.99, 1.0},
                yticklabels={96,97,98,99, 100},
                tick label style={font=\small}, label style={font=\small},
            ]
            \addplot[mark=oplus*, mark size=0.8pt, thick, color=rangeC4] coordinates { (80,1.0) (170,1.0) (260,1.0) (350,1.0) (440,0.9931506849315068) (530,1.0) (620,0.9805825242718447) (710,0.9788135593220338) (800,0.9812030075187969) (890,0.9831081081081081) };
            \addplot[mark=square*, mark size=0.8pt, thick, color=rangeB4] coordinates { (80,1.0) (170,1.0) (260,1.0) (350,1.0) (440,1.0) (530,0.9829545454545454) (620,0.9951456310679612) (710,0.9872881355932204) (800,0.9887218045112782) (890,0.9898648648648649) };
            \addplot[mark=diamond*, mark size=0.8pt, thick, dotted, color=rangeD3] coordinates { (80,1.0) (170,1.0) (260,1.0) (350,1.0) (440,0.9931506849315068) (530,0.9943181818181818) (620,0.9902912621359223) (710,0.9872881355932204) (800,0.9849624060150376) (890,0.9898648648648649) };
            \addplot[mark=star*, mark size=0.8pt, thick, dashdotted, color=rangeD4] coordinates { (80,1.0) (170,1.0) (260,1.0) (350,1.0) (440,1.0) (530,0.9943181818181818) (620,0.9902912621359223) (710,1.0) (800,0.9962406015037594) (890,0.9932432432432432) };
            \addplot[mark=Mercedes star*, mark size=0.8pt, thick, dashed, color=rangeD5] coordinates { (80,1.0) (170,1.0) (260,1.0) (350,1.0) (440,1.0) (530,0.9886363636363636) (620,0.9902912621359223) (710,0.9872881355932204) (800,0.9962406015037594) (890,0.9932432432432432) };
            \addplot[mark=+, mark size=0.8pt, thick, color=rangeD6] coordinates { (80,1.0) (170,1.0) (260,1.0) (350,1.0) (440,1.0) (530,1.0) (620,0.9951456310679612) (710,0.9957627118644068) (800,0.9962406015037594) (890,0.9966216216216216) };
            \addplot[mark=triangle, mark size=0.8pt, thick, dotted, color=rangeE3] coordinates { (80,1.0) (170,1.0) (260,1.0) (350,1.0) (440,1.0) (530,1.0) (620,0.9902912621359223) (710,1.0) (800,0.9962406015037594) (890,0.9966216216216216) };
            \addplot[mark=o, mark size=0.8pt, thick, dashdotted, color=rangeE4] coordinates { (80,1.0) (170,1.0) (260,1.0) (350,1.0) (440,1.0) (530,1.0) (620,0.9902912621359223) (710,1.0) (800,0.9962406015037594) (890,0.9932432432432432) };
            \addplot[mark=star, mark size=0.8pt, thick, dashed, color=rangeE5] coordinates { (80,1.0) (170,1.0) (260,1.0) (350,1.0) (440,1.0) (530,0.9886363636363636) (620,1.0) (710,0.9957627118644068) (800,0.9962406015037594) (890,0.9898648648648649) };
            \addplot[mark=Mercedes star, mark size=0.8pt, thick, color=rangeE6] coordinates { (80,1.0) (170,1.0) (260,1.0) (350,1.0) (440,1.0) (530,0.9717514124293786) (620,0.9741935483870968) (710,0.9661971830985916) (800,0.9962406015037594) (890,0.9932432432432432) };
            \addplot[mark=x, mark size=0.8pt, thick, color=exact] coordinates { (80,1.0) (170,1.0) (260,1.0) (350,0.3448275862068966) (440,0.0) (530,0.0) (620,0.0) (710,0.0) };
            \addplot[only marks, mark=x, mark options={scale=1.5, color=red, line width=1pt}] coordinates { (800,0.0) (890,0.0) };
            \end{axis}
        \end{tikzpicture}
    \end{minipage}\hfill
    \begin{minipage}{0.24\textwidth}
        \centering
        \begin{tikzpicture}
            \begin{axis}[
                width=5.5cm, height=5.5cm,
                title={\small Port utilization (\%)},
                grid=major, xmin=0, xmax=950, ymin=0, ymax=17, 
                xtick={260, 530, 890},
                ytick={0, 8, 16}, 
                tick label style={font=\small}, label style={font=\small},
            ]
            \addplot[mark=oplus*, mark size=0.8pt, thick, color=rangeC4] coordinates { (80,4.21) (170,6.46) (260,8.1) (350,9.61) (440,11.01) (530,12.0) (620,13.06) (710,13.8) (800,14.56) (890,15.22) };
            \addplot[mark=square*, mark size=0.8pt, thick, color=rangeB4] coordinates { (80,3.91) (170,5.93) (260,7.66) (350,8.96) (440,10.2) (530,11.18) (620,12.33) (710,13.07) (800,14.12) (890,14.69) };
            \addplot[mark=diamond*, mark size=0.8pt, thick, dotted, color=rangeD3] coordinates { (80,3.89) (170,5.92) (260,7.72) (350,8.91) (440,10.54) (530,11.52) (620,12.57) (710,13.12) (800,13.91) (890,14.67) };
            \addplot[mark=star*, mark size=0.8pt, thick, dashdotted, color=rangeD4] coordinates { (80,3.91) (170,5.99) (260,7.42) (350,9.16) (440,10.93) (530,11.77) (620,12.57) (710,14.01) (800,14.81) (890,15.49) };
            \addplot[mark=Mercedes star*, mark size=0.8pt, thick, dashed, color=rangeD5] coordinates { (80,4.02) (170,5.94) (260,7.72) (350,9.34) (440,10.57) (530,11.85) (620,12.54) (710,13.62) (800,13.98) (890,15.1) };
            \addplot[mark=+, mark size=0.8pt, thick, color=rangeD6] coordinates { (80,3.93) (170,5.84) (260,7.61) (350,9.26) (440,10.66) (530,11.55) (620,12.52) (710,13.89) (800,14.32) (890,14.95) };
            \addplot[mark=triangle, mark size=0.8pt, thick, dotted, color=rangeE3] coordinates { (80,4.02) (170,6.0) (260,7.72) (350,9.16) (440,10.57) (530,11.86) (620,12.87) (710,14.03) (800,14.67) (890,15.19) };
            \addplot[mark=o, mark size=0.8pt, thick, dashdotted, color=rangeE4] coordinates { (80,4.02) (170,6.0) (260,7.68) (350,9.2) (440,10.57) (530,12.04) (620,12.74) (710,13.69) (800,14.56) (890,14.9) };
            \addplot[mark=star, mark size=0.8pt, thick, dashed, color=rangeE5] coordinates { (80,4.02) (170,6.0) (260,7.72) (350,9.06) (440,10.33) (530,11.77) (620,12.88) (710,14.03) (800,14.36) (890,14.74) };
            \addplot[mark=Mercedes star, mark size=0.8pt, thick, color=rangeE6] coordinates { (80,4.02) (170,6.0) (260,7.72) (350,9.05) (440,10.57) (530,11.84) (620,13.17) (710,13.7) (800,14.05) (890,14.8) };
            \addplot[mark=x, mark size=0.8pt, thick, color=exact] coordinates { (80,4.02) (170,6.0) (260,7.56) (350,4.31) (440,5.22) (530,0.0) (620,0.0) (710,0.0) (800, 0.0) (890, 0.0) };
            \addplot[only marks, mark=x, mark options={scale=1.5, color=red, line width=1pt}] coordinates { (800,0.0) (890,0.0) };
            \end{axis}
        \end{tikzpicture}
    \end{minipage}\hfill
    \begin{minipage}{0.24\textwidth}
        \centering
        \begin{tikzpicture}
            \begin{axis}[
                width=5.5cm, height=5.5cm,
                title={\small Run-time (h)},
                grid=major, xmin=0, xmax=950, ymin=0, ymax=2.2, 
                xtick={260, 530, 890},
                ytick={0, 1, 2}, 
                tick label style={font=\small}, label style={font=\small},
            ]
            \addplot[mark=oplus*, mark size=0.8pt, thick, color=rangeC4] coordinates { (80,0.00546) (170,0.03038) (260,0.09348) (350,0.17379) (440,0.27901) (530,0.45961) (620,0.59607) (710,0.86124) (800,1.13603) (890,1.47702) };
            \addplot[mark=square*, mark size=0.8pt, thick, color=rangeB4] coordinates { (80,0.00121) (170,0.00487) (260,0.01419) (350,0.02346) (440,0.03587) (530,0.04881) (620,0.06736) (710,0.08869) (800,0.11108) (890,0.14778) };
            \addplot[mark=diamond*, mark size=0.8pt, thick, dotted, color=rangeD3] coordinates { (80,0.00125) (170,0.00622) (260,0.01785) (350,0.02351) (440,0.03539) (530,0.04881) (620,0.08014) (710,0.08886) (800,0.07347) (890,0.09483) };
            \addplot[mark=star*, mark size=0.8pt, thick, dashdotted, color=rangeD4] coordinates { (80,0.00129) (170,0.00550) (260,0.01371) (350,0.02084) (440,0.03398) (530,0.04369) (620,0.08014) (710,0.07872) (800,0.08873) (890,0.12403) };
            \addplot[mark=Mercedes star*, mark size=0.8pt, thick, dashed, color=rangeD5] coordinates { (80,0.00134) (170,0.00801) (260,0.01622) (350,0.03007) (440,0.03328) (530,0.06289) (620,0.08014) (710,0.08941) (800,0.17358) (890,0.13267) };
            \addplot[mark=+, mark size=0.8pt, thick, color=rangeD6] coordinates { (80,0.00135) (170,0.00758) (260,0.03845) (350,0.04267) (440,0.09442) (530,0.09119) (620,0.17029) (710,0.20261) (800,0.24258) (890,0.25697) };
            \addplot[mark=triangle, mark size=0.8pt, thick, dotted, color=rangeE3] coordinates { (80,0.00135) (170,0.00758) (260,0.03845) (350,0.04267) (440,0.09442) (530,0.18588) (620,0.22803) (710,0.27856) (800,0.40186) (890,0.37416) };
            \addplot[mark=o, mark size=0.8pt, thick, dashdotted, color=rangeE4] coordinates { (80,0.00135) (170,0.00758) (260,0.03845) (350,0.04267) (440,0.09442) (530,0.24097) (620,0.68005) (710,0.78049) (800,0.82873) (890,1.02311) };
            \addplot[mark=star, mark size=0.8pt, thick, dashed, color=rangeE5] coordinates { (80,0.00135) (170,0.00758) (260,0.03845) (350,0.04267) (440,0.09442) (530,0.50185) (620,0.66583) (710,0.73831) (800,1.25997) (890,1.46886) };
            \addplot[mark=Mercedes star, mark size=0.8pt, thick, color=rangeE6] coordinates { (80,0.00135) (170,0.00758) (260,0.03845) (350,0.04267) (440,0.09442) (530,0.84204) (620,1.16404) (710,1.13881) (800,1.16404) (890,1.47168) };
            \addplot[mark=x, mark size=0.8pt, thick, color=exact] coordinates { (80,0.00207) (170,0.04079) (260,0.15525) (350,2.04157) (440,2.05205) (530,2.09220) (620,2.11852) (710,2.16675) (800, 2.0) (890, 2.0)  };
            \addplot[only marks, mark=x, mark options={scale=1.5, color=red, line width=1pt}] coordinates { (800,2.0) (890,2.0) };
            \end{axis}
        \end{tikzpicture}
    \end{minipage}\hfill
    \centerline{Number of data streams requested}
    \centerline{\ref{mediumLegend}}
    \caption{Medium topology computational experiments}
    \label{fig:computational_results_medium}
\end{figure*}
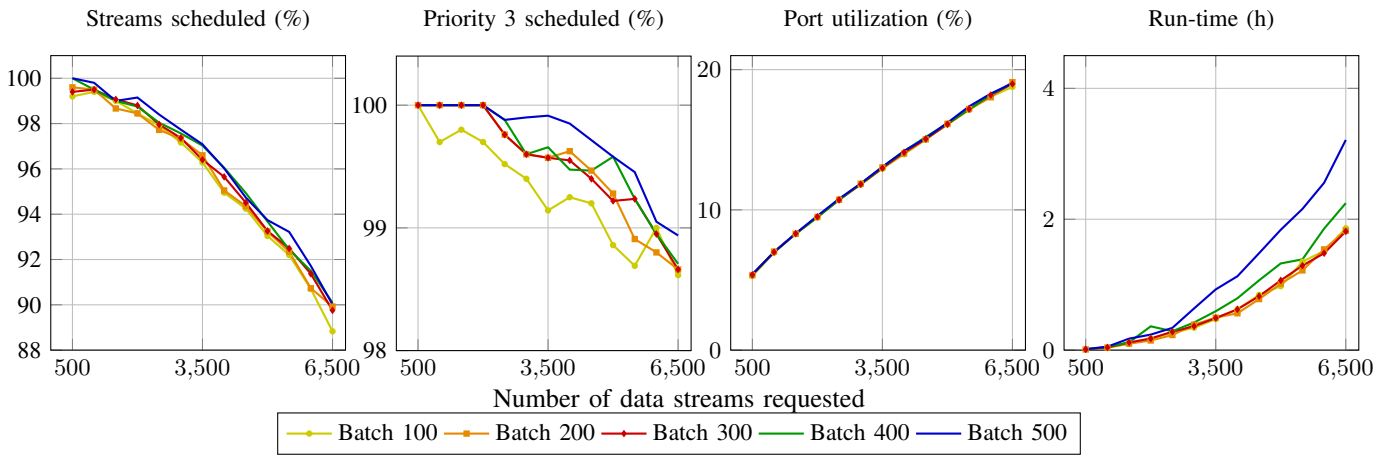
\begin{figure*}[!t]
    \centering
    \begin{minipage}{0.24\textwidth}
    \centering
    \begin{tikzpicture}
        \begin{axis}[
            width=5.5cm, height=5.5cm,
            title = {\small Streams scheduled (\%)},
            legend to name=largeLegend,
            legend columns=5,
            legend style={font=\small},
            grid=major,
            ylabel={},
            xmin=0, xmax=6800,
            ymin=0.88, ymax=1.01,
            xtick={500, 3500, 6500},
            ytick={0.88, 0.90, 0.92, 0.94, 0.96, 0.98, 1.00},
            yticklabels={88, 90, 92, 94, 96, 98, 100},
            yticklabel style={/pgf/number format/fixed, /pgf/number format/precision=1},
            tick label style={font=\small},
            label style={font=\small},
        ]
        \addplot[mark=oplus*, mark size=0.8pt, thick, color=rangeA6] coordinates { (500,0.992) (1000,0.994) (1500,0.99) (2000,0.9845) (2500,0.9792) (3000,0.9716666666666667) (3500,0.9628571428571429) (4000,0.9495) (4500,0.9424444444444444) (5000,0.9304) (5500,0.922) (6000,0.907) (6500,0.8883076923076923) }; \addlegendentry{Batch 100}
        \addplot[mark=square*, mark size=0.8pt, thick, color=rangeB5] coordinates { (500,0.996) (1000,0.995) (1500,0.9866666666666667) (2000,0.9845) (2500,0.9772) (3000,0.9733333333333334) (3500,0.966) (4000,0.9505) (4500,0.9435555555555556) (5000,0.9324) (5500,0.9234545454545454) (6000,0.9073333333333333) (6500,0.8992307692307693) }; \addlegendentry{Batch 200}
        \addplot[mark=diamond*, mark size=0.8pt, thick, color=rangeC5] coordinates { (500,0.994) (1000,0.995) (1500,0.9906666666666667) (2000,0.988) (2500,0.9796) (3000,0.9736666666666667) (3500,0.964) (4000,0.9565) (4500,0.9451111111111111) (5000,0.9326) (5500,0.9249090909090909) (6000,0.9136666666666666) (6500,0.8975384615384615) }; \addlegendentry{Batch 300}
        \addplot[mark=star*, mark size=0.8pt, thick,  color=rangeD6] coordinates { (500,1.0) (1000,0.995) (1500,0.99) (2000,0.9875) (2500,0.9804) (3000,0.9756666666666667) (3500,0.9702857142857143) (4000,0.9605) (4500,0.9493333333333334) (5000,0.9368) (5500,0.9243636363636364) (6000,0.915) (6500,0.9009230769230769) }; \addlegendentry{Batch 400}
        \addplot[mark=Mercedes star*, mark size=0.8pt, thick, color=rangeE5] coordinates { (500,1.0) (1000,0.998) (1500,0.99) (2000,0.9915) (2500,0.984) (3000,0.9773333333333334) (3500,0.9708571428571429) (4000,0.96025) (4500,0.9473333333333334) (5000,0.9374) (5500,0.9321818181818182) (6000,0.9171666666666667) (6500,0.9001538461538462) }; \addlegendentry{Batch 500}
        \end{axis}
    \end{tikzpicture}
    \end{minipage}\hfill
    \begin{minipage}{0.24\textwidth}
        \centering
        \begin{tikzpicture}
            \begin{axis}[
                width=5.5cm, height=5.5cm,
                title = {\small Priority 3 scheduled (\%)},
                grid=major,
                xmin=0, xmax=6800, ymin=0.98, ymax=1.004,
                xtick={500, 3500, 6500}, 
                ytick={0.98, 0.99, 1.0},
                yticklabels={98, 99, 100},
                yticklabel style={/pgf/number format/fixed, /pgf/number format/precision=1},
                tick label style={font=\small}, label style={font=\small},
            ]
            \addplot[mark=oplus*, mark size=0.8pt, thick, color=rangeA6] coordinates { (500,1.0) (1000,0.996996996996997) (1500,0.998) (2000,0.996996996996997) (2500,0.9951980792316927) (3000,0.994) (3500,0.9914236706689536) (4000,0.9924981245311327) (4500,0.992) (5000,0.9885954381752701) (5500,0.9869067103109657) (6000,0.99) (6500,0.9861495844875346) };
            \addplot[mark=square*, mark size=0.8pt, thick, color=rangeB5] coordinates { (500,1.0) (1000,1.0) (1500,1.0) (2000,1.0) (2500,0.9975990396158463) (3000,0.996) (3500,0.9957118353344768) (4000,0.9962490622655664) (4500,0.9946666666666667) (5000,0.992797118847539) (5500,0.989088925259138) (6000,0.988) (6500,0.9866112650046168) };
            \addplot[mark=diamond*, mark size=0.8pt, thick, color=rangeC5] coordinates { (500,1.0) (1000,1.0) (1500,1.0) (2000,1.0) (2500,0.9975990396158463) (3000,0.996) (3500,0.9957118353344768) (4000,0.9954988747186797) (4500,0.994) (5000,0.9921968787515006) (5500,0.9923622476813966) (6000,0.9895) (6500,0.9866112650046168) };
            \addplot[mark=star*, mark size=0.8pt, thick, color=rangeD6] coordinates { (500,1.0) (1000,1.0) (1500,1.0) (2000,1.0) (2500,0.9987995198079231) (3000,0.996) (3500,0.9965694682675815) (4000,0.994748687171793) (4500,0.9946666666666667) (5000,0.9957983193277311) (5500,0.9923622476813966) (6000,0.9895) (6500,0.987072945521699) };
            \addplot[mark=Mercedes star*, mark size=0.8pt, thick, color=rangeE5] coordinates { (500,1.0) (1000,1.0) (1500,1.0) (2000,1.0) (2500,0.9987995198079231) (3000,0.999) (3500,0.9991423670668954) (4000,0.9984996249062266) (5000,0.9957983193277311) (5500,0.994544462629569) (6000,0.9905) (6500,0.9893813481071099) };
            \end{axis}
        \end{tikzpicture}
    \end{minipage}\hfill
    \begin{minipage}{0.24\textwidth}
        \centering
        \begin{tikzpicture}
            \begin{axis}[
                width=5.5cm, height=5.5cm,
                title={\small Port utilization (\%)},
                grid=major,
                xmin=0, xmax=6800, ymin=0, ymax=21, 
                xtick={500, 3500, 6500},
                ytick={0, 10, 20}, 
                tick label style={font=\small}, label style={font=\small},
            ]
            \addplot[mark=oplus*, mark size=0.8pt, thick, color=rangeA6] coordinates { (500,5.28) (1000,6.95) (1500,8.28) (2000,9.46) (2500,10.75) (3000,11.79) (3500,12.94) (4000,13.97) (4500,15.02) (5000,16.08) (5500,17.13) (6000,18.07) (6500,18.77) };
            \addplot[mark=square*, mark size=0.8pt, thick, color=rangeB5] coordinates { (500,5.35) (1000,7.03) (1500,8.3) (2000,9.49) (2500,10.72) (3000,11.87) (3500,13.03) (4000,14.02) (4500,15.03) (5000,16.14) (5500,17.19) (6000,18.03) (6500,19.11) };
            \addplot[mark=diamond*, mark size=0.8pt, thick, color=rangeC5] coordinates { (500,5.37) (1000,6.99) (1500,8.31) (2000,9.51) (2500,10.71) (3000,11.83) (3500,12.98) (4000,14.1) (4500,15.06) (5000,16.11) (5500,17.19) (6000,18.19) (6500,18.98) };
            \addplot[mark=star*, mark size=0.8pt, thick, color=rangeD6] coordinates { (500,5.41) (1000,6.99) (1500,8.3) (2000,9.48) (2500,10.72) (3000,11.86) (3500,13.06) (4000,14.18) (4500,15.2) (5000,16.17) (5500,17.17) (6000,18.21) (6500,19.06) };
            \addplot[mark=Mercedes star*, mark size=0.8pt, thick, color=rangeE5] coordinates { (500,5.41) (1000,6.99) (1500,8.32) (2000,9.57) (2500,10.79) (3000,11.88) (3500,13.07) (4000,14.2) (4500,15.12) (5000,16.21) (5500,17.38) (6000,18.29) (6500,19.03) };
            \end{axis}
        \end{tikzpicture}
    \end{minipage}\hfill
    \begin{minipage}{0.24\textwidth}
        \centering
        \begin{tikzpicture}
            \begin{axis}[
                width=5.5cm, height=5.5cm,
                title={\small Run-time (h)},
                grid=major,
                ylabel={},
                xmin=0, xmax=6800, ymin=0, ymax=4.5, 
                xtick={500, 3500, 6500},
                ytick={0, 2, 4}, 
                tick label style={font=\small}, label style={font=\small},
            ]
            \addplot[mark=oplus*, mark size=0.8pt, thick, color=rangeA6] coordinates { (500,0.0136) (1000,0.0391) (1500,0.0995) (2000,0.1787) (2500,0.2637) (3000,0.3418) (3500,0.4749) (4000,0.6234) (4500,0.8442) (5000,0.9772) (5500,1.3496) (6000,1.5155) (6500,1.8619) };
            \addplot[mark=square*, mark size=0.8pt, thick, color=rangeB5] coordinates { (500,0.0099) (1000,0.0386) (1500,0.0987) (2000,0.1467) (2500,0.2305) (3000,0.3893) (3500,0.4964) (4000,0.5626) (4500,0.7794) (5000,1.0160) (5500,1.2179) (6000,1.5394) (6500,1.8229) };
            \addplot[mark=diamond*, mark size=0.8pt, thick,  color=rangeC5] coordinates { (500,0.0121) (1000,0.0408) (1500,0.1174) (2000,0.1774) (2500,0.2818) (3000,0.3651) (3500,0.4913) (4000,0.6233) (4500,0.8241) (5000,1.0657) (5500,1.2914) (6000,1.4803) (6500,1.8138) };
            \addplot[mark=star*, mark size=0.8pt, thick,  color=rangeD6] coordinates { (500,0.0120) (1000,0.0500) (1500,0.1178) (2000,0.3636) (2500,0.2888) (3000,0.4223) (3500,0.5945) (4000,0.7924) (4500,1.0665) (5000,1.3248) (5500,1.3868) (6000,1.8565) (6500,2.2458) };
            \addplot[mark=Mercedes star*, mark size=0.8pt, thick, color=rangeE5] coordinates { (500,0.0143) (1000,0.0526) (1500,0.1753) (2000,0.2370) (2500,0.3401) (3000,0.6371) (3500,0.9279) (4000,1.1288) (4500,1.4812) (5000,1.8391) (5500,2.1573) (6000,2.5567) (6500,3.2113) };
            \end{axis}
        \end{tikzpicture}
    \end{minipage}
    \centerline{Number of data streams requested}
    \centerline{\ref{largeLegend}}
    \caption{Large topology computational experiments}
    \label{fig:computational_results_large}
    \vspace{-0.4cm}
\end{figure*}

We evaluate the heuristic and the exact ILP across small, medium, and large topologies (see Figure~\ref{fig:computational_results_small}--\ref{fig:computational_results_large}). The exact ILP is used as a baseline since it provides the optimal solution when tractable, allowing us to directly assess the quality and scalability of the heuristic. However, the ILP quickly becomes intractable as problem size grows, making it suitable only for small and moderate scenarios.

In the small topology experiments (Figure~\ref{fig:computational_results_small}), the exact ILP schedules all streams at low loads but struggles to find an optimal solution beyond 160 streams due to the time limit. The batch heuristic maintains efficiency as load increases.

For the medium topology (Figure~\ref{fig:computational_results_medium}), the ILP can schedule up to 260 streams before running out of memory or time, while the heuristic continues to scale, scheduling more streams as load increases. Larger batch sizes improve overall schedulability, but require more resources. Smaller batches yield higher port utilization by favoring longer, less-contended paths, while larger batches maximize total schedulability with shorter paths.

In the large topology (Figure~\ref{fig:computational_results_large}), we schedule up to 6500 streams. Only the heuristic is evaluated due to the intractability of the exact ILP. The heuristic consistently schedules over 88\% of requested streams, with nearly all high-priority streams admitted, and maintains a reasonable runtime of 2 hours for offline scheduling as the network scales. 

Overall, the sequential batch ILP heuristic is predictable and scalable, reliably scheduling high-priority traffic, while the exact ILP is only practical for small instances.
\subsection{Robustness Experiments}
\label{eval:robust}
To evaluate robustness against wireless link latency deviations, we use two real-world 5G-like datasets: URLLC (low, deterministic latencies) and Det6G (long-tail delays). For each, we test 173 (URLLC) and 139 (Det6G) streams, with each stream traversing one or two wireless links. For Det6G, we consider upper bounds (UB) at 100th, 99.99th, and 99.9th percentiles. The robustness parameter \gls{r_sp} is varied from 0 to 1, and scheduling is performed using the exact ILP.

Figure~\ref{fig:robustness_gamma_merged} shows the percentage of streams scheduled (left) and the success probability (right) as \gls{r_sp} increases. As robustness increases, fewer streams can be scheduled due to larger, more conservative time windows for wireless transmissions, which reduces network capacity. However, the success probability—i.e., the likelihood that all deadlines are met despite wireless delay variations—increases accordingly. This illustrates the fundamental trade-off: higher \gls{r_sp} prioritizes reliability, while lower \gls{r_sp} maximizes network utilization. For Det6G, only the 100th percentile UB with \gls{r_sp}=1 achieves true 100\% reliability, reflecting the impact of long-tail delay distributions.

\begin{figure}[htbp]
    \centering
    \begin{minipage}{0.47\linewidth}
    \centering
    \begin{tikzpicture}
        \begin{axis}[
            width=4.8cm, height=4.8cm,
            title={\small Streams scheduled (\%)},
            legend to name=robustLegend,
            legend columns=4,
            legend style={font=\small},
            grid=major,
            xmin=0, xmax=1,
            ymin=0, ymax=1.05,
            xtick={0.0, 0.2, 0.4, 0.6, 0.8, 1.0},
            ytick={0.0, 0.20, 0.40, 0.60, 0.80, 1.0},
            yticklabels={0, 20, 40, 60, 80, 100},
            yticklabel style={/pgf/number format/fixed, /pgf/number format/precision=0},
            tick label style={font=\small},
            label style={font=\small},
        ]
        \addplot[mark=triangle*, mark size=0.8pt, thick, color=rangeC5] coordinates {
            (0.0,1.0) (0.1,0.965) (0.2,0.931) (0.3,0.908) (0.4,0.861) (0.5,0.827) (0.6,0.821) (0.7,0.803) (0.8,0.763) (0.9,0.728) (1.0,0.676)
        }; \addlegendentry{\small URLLC}
        
        \addplot[mark=square*, mark size=0.8pt, thick, color=rangeE6] coordinates {
            (0.0,1.0) (0.1,0.784) (0.2,0.604) (0.3,0.475) (0.4,0.432) (0.5,0.353) (0.6,0.324) (0.7,0.259) (0.8,0.245) (0.9,0.187) (1.0,0.187)
        }; \addlegendentry{\small UB 100th}
        
        \addplot[mark=diamond*, mark size=0.8pt, thick, dashed, color=rangeE5] coordinates {
            (0.0,1.0) (0.1,0.899) (0.2,0.791) (0.3,0.705) (0.4,0.604) (0.5,0.532) (0.6,0.504) (0.7,0.468) (0.8,0.403) (0.9,0.410) (1.0,0.381)
        }; \addlegendentry{\small UB 99.99th}
        
        \addplot[mark=otimes*, mark size=0.8pt, thick, dotted, color=rangeE4] coordinates {
            (0.0,1.0) (0.1,0.957) (0.2,0.806) (0.3,0.705) (0.4,0.691) (0.5,0.612) (0.6,0.576) (0.7,0.540) (0.8,0.532) (0.9,0.496) (1.0,0.432)
        }; \addlegendentry{\small UB 99.9th}
        \end{axis}
    \end{tikzpicture}
    \end{minipage}\hfill
    \begin{minipage}{0.47\linewidth}
    \centering
    \begin{tikzpicture}
        \begin{axis}[
            width=4.8cm, height=4.8cm,
            title={\small Success probability (\%)},
            grid=major,
            xmin=0, xmax=1,
            ymin=0, ymax=1.05,
            xtick={0.0, 0.2, 0.4, 0.6, 0.8, 1.0},
            ytick={0.0, 0.2, 0.4, 0.6, 0.8, 1.0},
            yticklabels={0, 20, 40, 60, 80, 100},
            yticklabel style={/pgf/number format/fixed, /pgf/number format/precision=0},
            tick label style={font=\small},
            label style={font=\small},
        ]
        \addplot[mark=triangle*, mark size=0.8pt, thick, color=rangeC5] coordinates {
            (0.0,0.036) (0.1,0.035) (0.2,0.147) (0.3,0.282) (0.4,0.457) (0.5,0.654) (0.6,0.815) (0.7,0.876) (0.8,0.930) (0.9,0.969) (1.0,1.000)
        }; 
        
        \addplot[mark=square*, mark size=0.8pt, thick, color=rangeE6] coordinates {
            (0.0,0.000) (0.1,0.681) (0.2,0.897) (0.3,0.943) (0.4,0.996) (0.5,1.000) (0.6,1.000) (0.7,1.000) (0.8,1.000) (0.9,1.000) (1.0,1.000)
        };
        
        \addplot[mark=diamond*, mark size=0.8pt, thick, dashed, color=rangeE5] coordinates {
            (0.0,0.000) (0.1,0.151) (0.2,0.452) (0.3,0.788) (0.4,0.899) (0.5,0.958) (0.6,0.989) (0.7,0.999) (0.8,1.000) (0.9,1.000) (1.0,1.000)
        };
        
        \addplot[mark=otimes*, mark size=0.8pt, thick, dotted, color=rangeE4] coordinates {
            (0.0,0.000) (0.1,0.072) (0.2,0.218) (0.3,0.340) (0.4,0.486) (0.5,0.657) (0.6,0.814) (0.7,0.920) (0.8,0.978) (0.9,0.993) (1.0,1.000)
        };
        \end{axis}
    \end{tikzpicture}
    \end{minipage}
    \vspace{0.05cm}
    \centerline{Robustness level \gls{r_sp}}
    \centerline{\ref{robustLegend}}
    \caption{Robustness: URLLC (173 streams), Det6G (139 streams)}
    \label{fig:robustness_gamma_merged}
    \vspace{-0.1cm}
\end{figure}
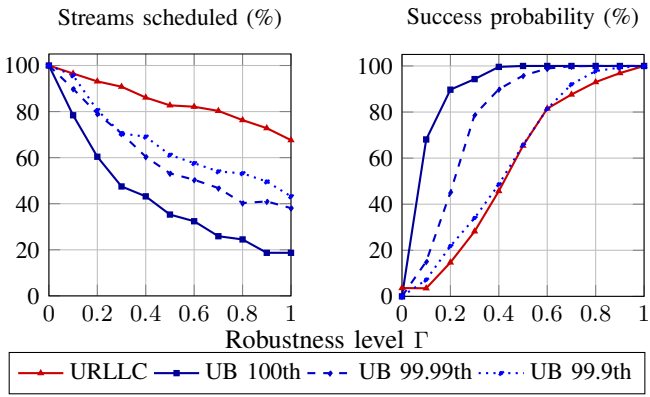

\section{Conclusion and Future Work}
This paper proposes a robust scheduling approach for TSN networks with wireless links, addressing latency uncertainties through a linear programming model and a scalable batch-scheduling ILP heuristic. The heuristic performs well for a moderate network load, with tunable batch size and robustness parameters that balance schedulability and reliability. In future work, we plan to explore clustering-based batch strategies, inclusion of best-effort traffic, cyclic stream dependencies, and stream-specific robustness based on traffic criticality.

\bibliographystyle{IEEEtran} 
\bibliography{bibliography} 

\begin{thebibliography}{10}
\providecommand{\url}[1]{#1}
\csname url@samestyle\endcsname
\providecommand{\newblock}{\relax}
\providecommand{\bibinfo}[2]{#2}
\providecommand{\BIBentrySTDinterwordspacing}{\spaceskip=0pt\relax}
\providecommand{\BIBentryALTinterwordstretchfactor}{4}
\providecommand{\BIBentryALTinterwordspacing}{\spaceskip=\fontdimen2\font plus
\BIBentryALTinterwordstretchfactor\fontdimen3\font minus \fontdimen4\font\relax}
\providecommand{\BIBforeignlanguage}[2]{{%
\expandafter\ifx\csname l@#1\endcsname\relax
\typeout{** WARNING: IEEEtran.bst: No hyphenation pattern has been}%
\typeout{** loaded for the language `#1'. Using the pattern for}%
\typeout{** the default language instead.}%
\else
\language=\csname l@#1\endcsname
\fi
#2}}
\providecommand{\BIBdecl}{\relax}
\BIBdecl

\bibitem{tsntaskgroup}
\BIBentryALTinterwordspacing
 {Time-Sensitive Networking (TSN) Task Group}. [Online]. Available: \url{https://1.ieee802.org/tsn/}
\BIBentrySTDinterwordspacing

\bibitem{zanbouri2023comprehensive}
K.~Zanbouri, M.~Noor-A-Rahim, J.~John, C.~J. Sreenan, H.~V. Poor, and D.~Pesch, ``{A Comprehensive Survey of Wireless Time-Sensitive Networking (TSN): Architecture, Technologies, Applications, and Open Issues},'' \emph{IEEE Communications Surveys \& Tutorials}, vol.~27, no.~4, pp. 2129--2155, 2025.

\bibitem{3gppURLLC}
\BIBentryALTinterwordspacing
{3GPP}, ``{Study on scenarios and requirements for next generation access technologies (Release 18)},'' Tech. Rep. 38.913, 2024. [Online]. Available: \url{https://portal.3gpp.org/desktopmodules/Specifications/SpecificationDetails.aspx?specificationId=2996}
\BIBentrySTDinterwordspacing

\bibitem{kaynak2024}
{\"O}.~O. Kaynak, A.~Kassler, A.~Fischer, O.~Dobrijevic, and H.~Chahed, ``{TSN Scheduling Robust to Wireless Performance Uncertainties: A Problem and Model Definition},'' in \emph{Proc. of the KuVS Fachgespr{\"a}ch - W{\"u}rzburg Workshop on 6G Networks (WueWoWAS'24)}, 2024, p.~4.

\bibitem{Craciunas2016}
S.~S. Craciunas, R.~S. Oliver, M.~Chmel\'{\i}k, and W.~Steiner, ``{Scheduling Real-Time Communication in IEEE 802.1Qbv Time Sensitive Networks},'' in \emph{Proc. of the 24th Int'l Conference on Real-Time Networks and Systems}, 2016, pp. 183--192.

\bibitem{Durr2016}
F.~D\"{u}rr and N.~G. Nayak, ``{No-wait Packet Scheduling for IEEE Time-sensitive Networks (TSN)},'' in \emph{Proc. of the 24th Int'l Conference on Real-Time Networks and Systems}, 2016, pp. 203--212.

\bibitem{egger2025end}
S.~Egger, J.~Gross, J.~Sachs, G.~P. Sharma, C.~Becker, and F.~Dürr, ``{End-to-End Reliability in Wireless IEEE 802.1Qbv Time-Sensitive Networks},'' in \emph{2025 IEEE/ACM 33rd International Symposium on Quality of Service (IWQoS)}, 2025, pp. 1--10.

\bibitem{Ginthor2020}
D.~Ginthör, R.~Guillaume, J.~von Hoyningen-Huene, M.~Schüngel, and H.~D. Schotten, ``{End-to-end Optimized Joint Scheduling of Converged Wireless and Wired Time-Sensitive Networks},'' in \emph{2020 25th IEEE International Conference on Emerging Technologies and Factory Automation (ETFA)}, 2020, pp. 222--229.

\bibitem{Ginthor2021}
D.~Ginthör, R.~Guillaume, M.~Schüngel, and H.~D. Schotten, ``{Robust End-to-End Schedules for Wireless Time-Sensitive Networks under Correlated Large-scale Fading},'' in \emph{2021 17th IEEE International Conference on Factory Communication Systems (WFCS)}, 2021, pp. 115--122.

\bibitem{Sharma2024}
G.~P. Sharma, W.~Tavernier, D.~Colle, M.~Pickavet, J.~Haxhibeqiri, J.~Hoebeke, and I.~Moerman, ``{End-to-End No-wait Scheduling for Time-Triggered Streams in Mixed Wired-Wireless Networks},'' \emph{Journal of Network and Systems Management}, vol.~32, no.~3, p.~65, 2024.

\bibitem{lizhong2024}
Z.~Li, J.~Yang, C.~Guo, J.~Xiao, T.~Tao, and C.~Li, ``{A Joint Scheduling Scheme for WiFi Access TSN},'' \emph{Sensors}, vol.~24, no.~8, 2024.

\bibitem{bertsimas}
D.~Bertsimas and M.~Sim, ``{The Price of Robustness},'' \emph{Operations Research}, vol.~52, no.~1, pp. 35--53, 2004.

\bibitem{Stüber2023}
T.~Stüber, L.~Osswald, S.~Lindner, and M.~Menth, ``{A Survey of Scheduling Algorithms for the Time-Aware Shaper in Time-Sensitive Networking (TSN)},'' \emph{IEEE Access}, vol.~11, pp. 61\,192--61\,233, 2023.

\bibitem{Dinand2023patent}
R.~Dinand, G.~Eriksson, K.~Wang, M.~Matti, and J.~Jeong, ``{Communication System with De-jitter Buffer for Reducing Jitter},'' US patent US 11,765,094 B2, Sep. 19, 2023.

\bibitem{yens}
J.~Y. Yen, ``{Finding the K Shortest Loopless Paths in a Network},'' \emph{Management Science}, vol.~17, no.~11, pp. 712--716, 1971.

\bibitem{det6gdata}
G.~P. Sharma, D.~Patel, J.~Sachs, M.~De~Andrade, J.~Farkas, J.~Harmatos, B.~Varga, H.-P. Bernhard, R.~Muzaffar, M.~Ahmed, F.~Dürr, D.~Bruckner, E.~M. De~Oca, D.~Houatra, H.~Zhang, and J.~Gross, ``{Toward Deterministic Communications in 6G Networks: State of the Art, Open Challenges and the Way Forward},'' \emph{IEEE Access}, vol.~11, pp. 106\,898--106\,923, 2023.

\bibitem{URLLC_experiment}
P.~Kehl, J.~Ansari, M.~H. Jafari, P.~Becker, J.~Sachs, N.~König, A.~Göppert, and R.~H. Schmitt, ``{Prototype of 5G Integrated with TSN for Edge-Controlled Mobile Robotics},'' \emph{Electronics}, vol.~11, no.~11, 2022.

\end{thebibliography}
\end{document}